\documentclass[preprint,3p,12pt,numbers,sort&compress]{elsarticle}
\usepackage{lipsum}
\usepackage{nomencl}
\makenomenclature
\makeatletter
\def\ps@pprintTitle{%
 \let\@oddhead\@empty
 \let\@evenhead\@empty
 \def\@oddfoot{}%
 \let\@evenfoot\@oddfoot}
\makeatother
\usepackage{tcolorbox}
\usepackage[section]{placeins}
\usepackage[shortlabels]{enumitem}
\usepackage{amsmath,amsthm,amssymb,amsfonts}
\usepackage{graphicx,float}
\usepackage[colorinlistoftodos]{todonotes}
\usepackage[colorlinks=true, allcolors=blue]{hyperref}
\usepackage{amsfonts}
\usepackage{lscape}
\usepackage{rotating}
\usepackage{color}
\usepackage{soul}
\usepackage{subcaption}
\usepackage{natbib}
\usepackage[symbol]{footmisc}
\usepackage{ulem}
\usepackage{miller}
\usepackage{mathtools}

\newcommand {\eqn}[1] {Eq.~(\ref{#1})}

\newcommand {\fig}[1] {Fig.~\ref{#1}}

\makeatletter
\newcommand{\mathleft}{\@fleqntrue\@mathmargin0pt}
\newcommand{\mathcenter}{\@fleqnfalse}
\makeatother
\makeatletter
\newcommand*\bigcdot{\mathpalette\bigcdot@{.5}}
\newcommand*\bigcdot@[2]{\mathbin{\vcenter{\hbox{\scalebox{#2}{$\m@th#1\bullet$}}}}}
\makeatother
\makeatletter
\AtBeginDocument{%
  \expandafter\renewcommand\expandafter\subsection\expandafter{%
    \expandafter\@fb@secFB\subsection
  }%
}
\makeatother
\renewcommand{\thefootnote}{\fnsymbol{footnote}}
\begin{document}
\begin{frontmatter}
\title{An Experimentally Informed Continuum Grain Boundary Model}
 %Department of Mechanical Engineering, IIT Bombay}
\author[IITB]{S.Syed Ansari \corref{cor1}}%\corref{cor1}}
\ead{syed.ansari@iitb.ac.in}
\address[IITB]{Department of Mechanical Engineering, Indian Institute of Technology Bombay, Powai, Mumbai--400076, India}
\author[CMU]{Amit Acharya\corref{cor2}}%\corref{cor2}}
\ead{acharyaamit@cmu.edu}
\address[CMU]{Department of Civil \& Environmental Engineering, and   Center for Nonlinear Analysis, Carnegie Mellon University, Pittsburgh, PA 15213, USA}
\author[IITB]{Alankar Alankar\corref{cor3}}
\ead{alankar.alankar@iitb.ac.in}
\cortext[cor3]{Corresponding author, Tel.: +91-9769415356, Fax: +91-22-25726875}
% \cortext[cor3]{Tel.: +91-9769415356}
% \cortext[cor3]{Fax: +91-22-25726875}
 %\author[]{Name2 Surname3}
 %\hrulefill
\begin{comment}
\begin{highlights}
\item The presented continuum grain boundary model incorporates experimentally measured grain boundary energy data as a function of misorientation to simulate various idealized grain boundary evolutions and their equilibria (in 1--D). To our knowledge, while natural, this is novel, and results in significantly different energetic characteristics than typically considered in theoretical (and simulation-based) studies of grain boundary behavior.
\item A computational scheme is presented to handle the constraints arising from the presence of kinks in orientation profiles and the strong non-convexity of the experimentally measured energies.
\item The present model is demonstrated using two energy density functions, namely, a smooth energy density (SED) and cusp energy density (CED). 
\item Various evolutions and their equilibria (in 1--D) recover idealized features of real physical systems such as equilibrium high--angle grain boundaries (HAGBs), grain rotation, grain growth, heavily deformed (static) microstructure often observed after the deformation process, and strong metastability. All the transition layers represent dislocation walls, and many of the equilibria resemble polygonized domains.
\end{highlights}
\end{comment}
%
\begin{abstract}
A continuum grain boundary model is developed that uses experimentally measured grain boundary energy data as a function of misorientation to simulate idealized grain boundary evolution in a 1--D grain array. The model uses a continuum representation of the misorientation in terms of spatial gradients of the orientation as a fundamental field. The grain boundary energy density employed is non--convex in this orientation gradient, based on physical grounds. Simple gradient descent dynamics of the energy are utilized for idealized microstructure evolution, which requires higher--order regularization of the energy density for the model to be well--set; the regularization is physically justified. Microstructure evolution is presented using two plausible energy density functions, both defined from the same experimental data: a `smooth' and a `cusp' energy density. Results of grain boundary equilibria and microstructure evolution representing grain reorientation in one space dimension are presented. The different shapes of the energy density functions representing a common data set are shown to result in different overall microstructural evolution of the system. Mathematically, the constructed energy functional formally is of the Aviles--Giga/Cross--Newell type but with unequal well--depths, resulting in a difference in the structural feature of solutions that can be identified with grain boundaries, as well as in the approach to equilibria from identical initial conditions. This study also investigates the metastability of grain boundaries. It supports the general thermodynamics belief that they persist for extended periods before eventually vanishing due to the lowest energy configuration favored by fluctuations over infinite time.
\end{abstract}
\end{frontmatter}
%
%\hrulefill
%\hrulefill
%\begin{enumerate}[start=1,label={\bfseries Q \arabic*.}]
\section{Introduction}\label{section:intro}
\noindent Microstructure is the primary factor in deciding the bulk properties of polycrystalline materials \cite{sutton2006interfaces}. The understanding of grain boundaries plays a vital role in assessing  overall polycrystalline microstructural evolution and equilibria \cite{kobayashi2000continuum,zhang2021equation,chen2020temperature,wei2019continuum,ROLLETT19891227,SAROCHAWIKASIT2021101186,Ratanaphan2019,cahn1958free,MARTINELABOISSONIERE2019100280,ESEDOLU2016209}. This work develops a grain boundary model at the continuum scale, informed by experimentally measured grain boundary energy. 

It is a simple, but significant, physical fact about solids (and, generally, most phases of condensed matter free from electromagnetic effects) that a superposed rigid rotation of a body does not incur an energy cost. Thus, in any local description of the energy density of a solid, a rotation (or an orientation w.r.t some reference frame) cannot affect the energy density, as all rotations are energetically equivalent. Rotation gradients can be ascribed such a cost. Grain boundaries are narrow transition regions of gradients in orientation. Based on an understanding of the structure and solutions of Allen-Cahn equations, while it is tempting to fit the modeling of grain boundaries into such a scheme employing a non-convex energy density in orientations with gradients penalized accordingly, the rotational invariance mentioned above makes the idea physically unpalatable. On the other hand, the standard idealization of viewing a collection of grains joined by a grain boundary network as kinematically described by an orientation field can and, arguably, should be, thought of as an elastic medium with defects, with the distortion field on the body constrained to take values in the set of all proper orthogonal tensors. The typical energy density of a local elastic medium is rotationally invariant disallowing any nontrivial dependence on the pointwise orientation, but allowing a higher order dependence on the rotation gradient, work-conjugate to volumetric couple stresses, and resulting torques per unit area on surfaces. Such an approach also has the natural advantage of allowing a seamless conceptual generalization of coupling the mechanics of grain boundaries to mechanical stress and applied loads when the ignored elastic strain is accounted for instead of only rotation, along with the defect dynamics mediating such coupling. It is such a model that we pursue here, in its simplest possible realization in 1 space dimension and time, for the abovementioned reasons of physical appropriateness and holistic, subsequent generalization.

\begin{figure}[H]
\centering
% [trim={left bottom right top},clip]
%\includegraphics[trim={0 7cm 0 7cm}, clip, width=0.75\linewidth]{Fig1}
\includegraphics[width=0.75\linewidth]{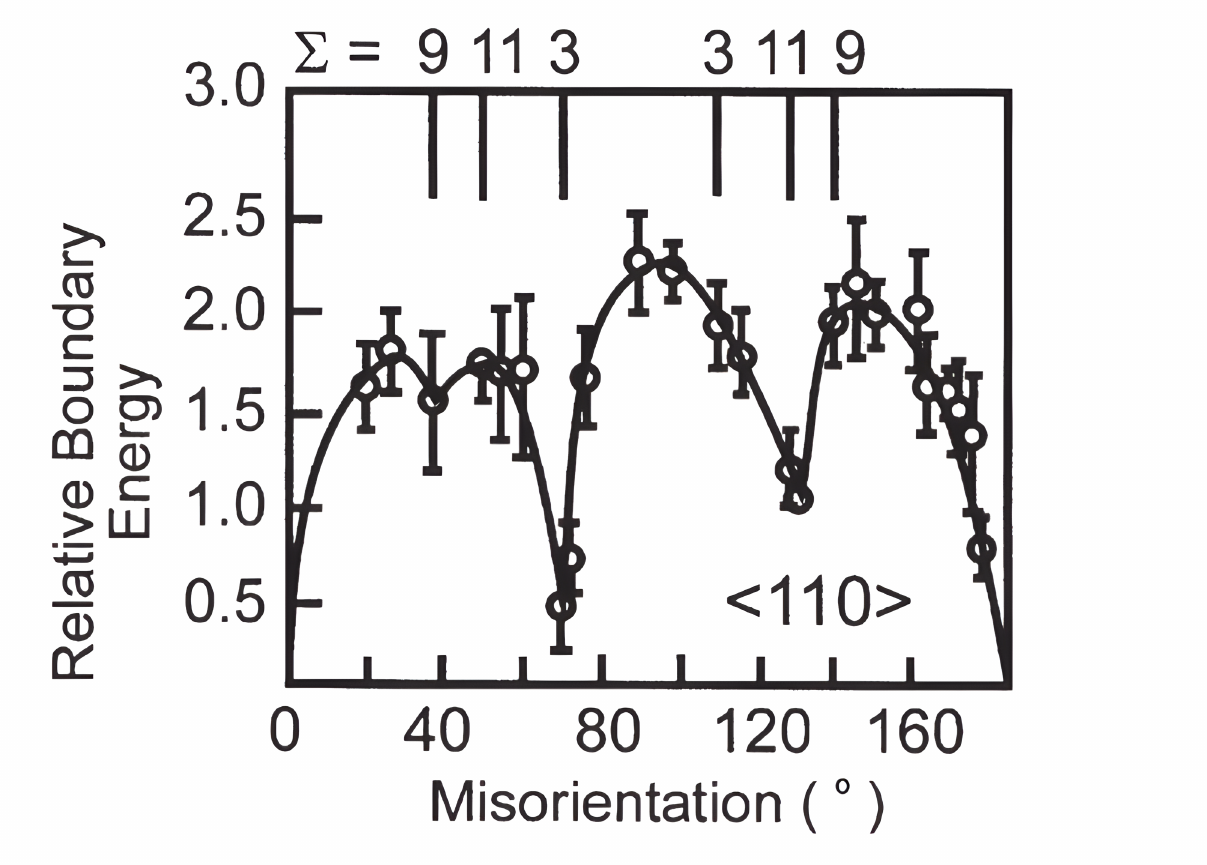}
\caption{Energies at 650$^o$C for symmetrical \hkl<100> and \hkl<110> tilt boundaries in aluminum originally reported in Hasson and Goux \cite{hasson1971interfacial}. Reprinted from Recrystallization and Related Annealing Phenomena, Second Edition, F.J. Humphreys, M. Hatherly, The Structure and Energy of Grain Boundaries, Page No.102, Copyright (2004), with permission from Elsevier \cite{HUMPHREYS200491}.}
\label{hassongouxtilt}
\end{figure}

The fundamental ingredient of the proposed model is a grain boundary energy density function that is directly inferred from experimental data reported in the literature as a function of misorientation ($\Delta \theta$). A model parameter $l$ is invoked that represents a physically measurable grain boundary width, and it is assumed that all measured misorientations $\Delta \theta$ occur over this width. With this direct correspondence, the measured grain boundary energy density variations as a function of misorientation (see, e.g., \fig{hassongouxtilt}) are converted to functions of orientation gradients ($\nabla \theta$) as follows
\begin{equation}
\label{misor_conv}
\Delta \theta =l \nabla \theta.
\end{equation}
Since measured grain boundary energy densities are strongly non--convex, representing the important physical fact that grain boundaries occur at special misorientations, our grain boundary energy density is naturally so, as well. At this point in time, we adopt a gradient flow of the energy functional based on the energy density to describe simple (local) energy minimizing dynamics; as is well--understood, a non--convex energy density in orientation gradients results in a gradient flow dynamics that cannot possess continuous dependence w.r.t. initial data on orientations, a feature arising from the dynamics of the backward heat equation (locally) \mbox{\cite{strang1986appliedmath}}. Consequently, we regularize with a second--orientation gradient energy penalty to obtain an energy functional given by
\begin{equation} \label{Overall Free Energy of the system}
\textit{F}=\int \left[{\phi (|\nabla \theta|) + \epsilon^2|\nabla^2\theta|^2}\right] dv, 
\end{equation}
where $\phi$ is the energy density function designed from experimental data, and the other term in the functional being the energy penalty to sharp kinks in the orientation profile (that also provides a robust principal part for the corresponding gradient flow evolution equation). We physically justify the regularization.
\par Our model looks formally similar to the Aviles--Giga \cite{aviles1996distance} (A--G) and Cross--Newell \cite{cross1984convection} (C--N) functionals, if in \eqn{Overall Free Energy of the system}, the $\phi$ is replaced by $\phi_{AG} = \left({1-|\nabla\theta|^2}\right)^2$, an energy density with equal--depth wells (and in C--N the regularizer would be $(\nabla \cdot \nabla \theta)^2)$. The A--G and C--N models have been extensively studied in the mathematical literature, see, e.g., \cite{kohn2007energy,jabin2002line,ercolani2009variational,glasner2006grain}. One difference between the typical A-G functional that have been studied and our model is that ours physically requires unequal depths to endow grain boundaries with non--vanishing energy content. In the standard interpretation of the A--G model, the troughs of the energy density correspond to the phases (i.e., regions of slope $\pm 1$, the `grains') and the kinks as the phase or grain boundaries (in 1--D).
%\\
\par
\begin{figure}%%[H]
\centering
\includegraphics[width=0.75\linewidth]{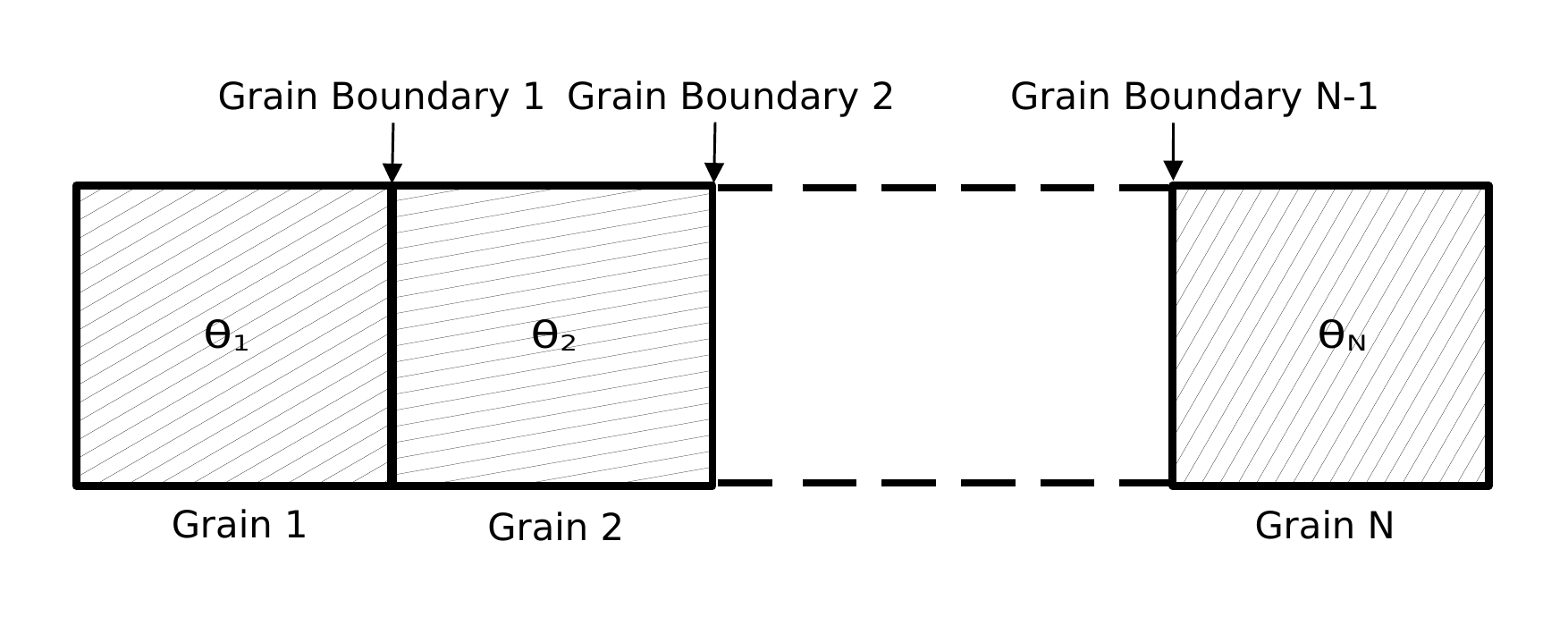}
\caption{A model bar represents a 1--D array of grains separated by straight vertical grain boundaries. Patterns inside each grain represent its orientation.}
\label{Bar with 3 Grains}
\end{figure}

 In our model, the grain boundary is simply a phase, as are the grains (of different spatial extent). We show numerically that the dynamics of the model reflects qualitatively expected behavior. We are  not aware of a gradient flow of the A--G, C--N type models in 1 space dimension that has been shown to represent phase boundary motion (without any further coupling to other fields). In our case, this is realistic as, for a simplistic model allowing spatial variations in only one space dimension, only straight physical grain boundaries can be represented that extend from one boundary of the domain to another as shown in \fig{Bar with 3 Grains}, and it is experimentally observed that straight (as opposed to curved) boundaries in bi--crystal configurations (involving boundaries without constraints of multi--junctions) do not move in the absence of stress or electromagnetic fields \cite{GOTTSTEIN20019,molodov2015,KIRCH2007939,KIRCH20084998}.

We also mention that in the elasticity of compatible phase transformations, this can be an appropriate mathematical model as well \cite{kohn1994surface}, with an important physical difference in interpretation: there the phases correspond to constant deformation/displacement gradients, whereas in our model, the `bulk' phases (i.e., excluding the narrow phase boundary `phases') correspond to constant rotation gradients, with the result that the work--conjugate quantities there (in the context of such a mathematical model) are forces and stresses, whereas in our case it is moments and couple--stresses.
Kobayashi, Warren, and Carter \cite{kobayashi2000continuum} developed a two--dimensional (2--D) continuum grain boundary (KWC) model using two variables $\nabla \theta$ and $\eta$ (degree of crystallinity) to simulate grain growth by curvature--driven grain shrinkage and grain rotation mechanisms. It has been extended to three dimensions (3--D) where the grain boundary energy is defined on a five--dimensional misorientation, and inclination space \cite{admal2019three}. According to the viewpoint put forward in \mbox{\cite{kobayashi2000continuum}}, the earlier WCK model \mbox{\cite{warren1998phase}}
is not adapted to handle non--convex grain boundary energy density functions for the reason that localized transition layers cannot be predicted as global minimizers of their energy. However, the type of model energy density function studied in WCK \mbox{\cite{warren1998phase}} is arguably, the most natural way of representing the entire spectrum of measured grain boundary energies in polycrystalline materials. An important philosophical difference between our work (with an energy density component similar in spirit to that in \mbox{\cite{warren1998phase}}), and the subsequent KWC and KWCL \mbox{\cite{Lobkovsky2001}} model is that we do not insist on single transition layer grain boundary solutions to be global minima of our energy functional and our simulations - which require a somewhat careful numerical treatment of the term involving the non--convex energy density in the governing gradient--descent equation - show that such solutions, corresponding to local minima of the functional, persist over significant times and are stable to physically natural high wave--number, small amplitude perturbations as shown in \mbox{\ref{appendix:a}}.

In general, our experimental grain boundary energy density--informed model displays a high level of metastability in its solutions that may be physically interpreted as grain boundary equilibria.

An outline of the paper is as follows: A summary of essential notation is followed by a brief literature review in Section \ref{section:brief_literature}, and the mathematical development of the model is discussed in Section \ref{sec:cgb_model}. Section \ref{section:numericalimplementation} discusses the numerical implementation of the model. Section \ref{section:results_discussion} discusses the simulations of microstructure evolution, including grain growth by grain rotation in 1--D. Evolution of a `random' initial condition to an equilibrium 1--D array of grains is also discussed in Section \ref{section:results_discussion}, with resemblance to an idealized grain microstructure with high--angle grain boundaries and some grains containing a high concentration of sub-boundaries. Concluding remarks are presented in Section \ref{section:conclusion}.
\section*{Notation}
\noindent
\begin{equation*}
\begin{aligned}
F & \qquad \mbox{Total free energy of the system}\\
\varphi & \qquad \mbox{Overall free energy density}\\
l & \qquad \mbox{Typical width of a grain boundary}\\
M & \qquad \mbox{Grain boundary mobility}\\
\theta & \qquad \mbox{Grain orientation}\\
\phi_{c} & \qquad \mbox{Cusp energy density (CED) function}\\
\phi_{s} & \qquad \mbox{Smooth energy density (SED) function}\\
\Delta \theta & \qquad \mbox{Misorientation between two grains}\\
L & \qquad \mbox{Total length of the domain}\\
k & \qquad \mbox{Wave number}\\
\partial_x \theta & \qquad \mbox{Orientation gradient in one space dimension}\\
\epsilon & \qquad \mbox{Material constant penalizing sharp kinks in orientation}
\end{aligned}
\end{equation*}
\subsection{A brief survey of relevant literature}\label{section:brief_literature}
\noindent Phase--field models have helped simulate microstructural evolution \cite{Protavas_PFMMSE_2010,Bhadeshia_PFM2010,Chen_PFM_AR_2002,MOELANS2008268}. A detailed review of various phase--field models is available in \cite{TOURRET2022100810}. Grain growth is simulated in \cite{FAN1997611,KRILLIII20023059,ZHANG2020211,MCKENNA2014125,JOHNSON2014134, Morin1995} using Allen--Cahn \cite{allen1979microscopic} type equations that use a non--conserved order parameter and a non--convex energy density function.

The effect of triple junctions and grain boundary thickness on grain growth have been studied using the Allen--Cahn type phase--field model in \cite{McKenna2009}. The aforementioned models \cite{FAN1997611,KRILLIII20023059,ZHANG2020211,MCKENNA2014125,JOHNSON2014134, Morin1995, McKenna2009, Kim_2014} employ orientation itself as a variable that enters the energy density, which is not consistent with frame indifference/rotational invariance of the energy density \cite{kobayashi2000continuum}. Allen--Cahn type equations are also believed not to be able to represent grain growth by the grain rotation mechanism \cite{kobayashi2000continuum}.

A variational approach is employed by Ta'asan and co--workers \cite{kinderlehrer2006} to compute curvature--driven grain growth based on Mullins' model \cite{Mullin1956} where the grain boundary energy density is a function of the grain boundary plane normal, along with the Herring condition \cite{herring1951physics, Herring1953} imposed at the triple junctions. Grain growth in a large system is simulated by Esedoglu and co-workers \cite{ELSEY20098015} based on mean curvature-driven motion while satisfying the Herring condition at triple junctions. Their algorithm uses the linear diffusion equation and signed distance functions \cite{ESEDOGLU20101017}.

Glasner computationally explored the gradient flow dynamics of the A--G functional \cite{glasner2006grain}, as well as derived a matched-asymptotic-expansion based model for the evolution of grain boundaries and junctions corresponding to the singular limit of the gradient flow of the A--G functional. Grain boundary motion in the limit model is found to arise from curvature of the interfaces and variation in the ``line energy'' along grain boundaries. Interestingly, based on the results of numerical simulations, Glasner speculates on the possibility of grain boundaries terminating in the bulk, in other words to the presence of disclinations. Oudet et al. \mbox{\cite{Oudet2011}} introduced a diffuse interface approximation of branched transport. The model has some similarities with ours in that it involves a concave energy density function ($|u|^\beta, 0< \beta <1$) corresponding to the solution variable $u$ (our non--linearity is in the gradient) and a regularisation term. The corresponding minima of the energy density is at  $u=0$ and because of a non-vanishing specified divergence constraint (or specified average constraint), large values of $|u|$ on small sets also have to occur, which may be thought of as another phase. Energy concentrates on the boundary between the phases corresponding to $u=0$ and $|u|=\infty$.
\par A crystal plasticity model combined with the continuum phase--field model (KWC) is used to simulate shear--induced grain boundary motion, grain boundary sliding, curvature--driven grain boundary rotation, and curvature--driven grain shrinkage in unison \cite{admal2018unified}. In the context of a polycrystal, 2--D \cite{basakgupta_2D_2014} and 3--D \cite{basakgupta_3D_2015} models by Basak and Gupta are notable in the sense that they consider diffusion controlled incoherent interfaces in the presence of junctions. Their framework is categorized as a \textit{sharp interface model} and employs curvature dependent GB energy and surface diffusion motivated by the work of Gurtin and co-workers \cite{gurtinjabbour_3D_2002}.

Efforts to deduce relationships between the evolution of grain boundaries/networks and geometric parameters describing them based on experimental data appear to be in a state of continual evolution. According to \cite{rollett2001grain}, the five-parameter grain boundary distribution and grain boundary mobility are strongly correlated. In contrast, it is reported in \cite{ZHANG2020211} that there is no correlation between the five-parameter grain boundary distribution and grain boundary mobility. According to \cite{Aditi_CMU_2021}, there is almost no correlation observed between curvature and grain boundary velocity, but a strong correlation is observed between the five-parameter grain boundary distribution and grain boundary velocity in FCC metals. Thus, it is perhaps fair to say that there is a need for a definitive model that mechanistically describes grain boundary network evolution and the experimental observations above while allowing for experimentally measured descriptions for the energies of grain boundaries.

Alternative to the curvature--driven models, Srolovitz and co--workers presented a continuum equation of motion for a grain boundary and corresponding models based on the discrete disconnection mechanism in a series of recent works \cite{Srolovitz_GBE_2017, wei2019continuum, HAN2021117178} and extended it to the polycrystals \cite{zhang2021equation}.

To our knowledge, there are no existing phase--field models that use energy densities that are rotationally invariant (or frame-indifferent) and use experimentally measured or atomistically determined grain boundary energies to simulate the dynamics of different types of grain boundaries. Further, the experimentally measured grain boundary energy density of aluminum symmetrical tilt boundaries shown in \fig{hassongouxtilt} indicates that the equilibrium high--angle grain boundaries (HAGBs) corresponding to troughs have lower energy than certain low--angle grain boundaries (LAGBs) that explain the existence of equilibrium HAGBs. We are unaware of any rotationally invariant phase--field based grain boundary models that account for this aspect. Considering these points, the main objectives of the current work is to develop a model:
 
\begin{itemize}
    \item in which an experimentally measured GB energy function drives the evolution of grain boundaries,
    \item that addresses rotational invariance,
    \item that captures equilibrium transition layers in a 1--D grain array as an idealized representation of equilibrium HAGBs,
    \item is built upon a robust computational scheme to handle the constraints arising from the presence of kinks in orientation profiles and the strong non-convexity of the experimentally measured energies,
    \item that recovers some aspects of idealized grain rotation and grain growth in 1--D.
    
\end{itemize}
The model presented in this work can capture equilibrium transition layers in a 1--D grain array as an idealized representation of equilibrium HAGBs. Occasionally, it also produces a wide equilibrium transition layer representative of grains with a large number of subgrain boundaries. Our model is able to address the grain growth mechanism by grain reorientation (idealized to be along one space dimension). Our work is complementary to \cite{kinderlehrer2006} in that their grain growth model considers the grain boundaries and triple junctions as discrete objects, with additional rules for topological changes in the grain boundary network, whereas we view such features as localized features of a continuous field of orientation. The possibility of the energy density to be a function of the misorientation is mentioned in \cite{kinderlehrer2006}, with the bulk of the model development and implementation ignoring this dependence, focusing instead on the dependence of the energy on the boundary normal alone in 2--D. Our current work is in 1--D (so the boundary normal is constant) and we focus on the dependence of the energy density on the misorientation.

Our model can be placed in the context of the continuum mechanics of dislocation and g.disclination defects \cite{Acharya2012,Acharya2015,ZHANG2018Acharya,ZHANG2018finite,hirth2021straight}. In this setting, grain/phase boundaries are represented by (third--order tensor) eigenwall fields, $S$, concentrated on the physical grain boundary regions, and their terminations represent g.disclinations - triple junctions, e.g., represent the superposition  of three g.disclination curves (for an example related to a penta--twin, see \cite{ZHANG2018finite}). An evolution equation, determined essentially from the statement of conservation of g.disclination topological charge $\mathnormal{\Pi}$, for the eigenwall field is given by (using `small deformation' kinematics for the sake of conveying ideas)
\begin{equation*}
\begin{aligned}
\mathnormal{\Pi} & = - \mbox{curl}\, S; & \mathnormal{\Pi}_{ijk} = - \varepsilon_{krs} \partial_r S_{ijs} \nonumber\\
\partial_t \mathnormal{\Pi} & = - \mbox{curl}\,\left(\mathnormal{\Pi} \times V^{(S)} \right); & \partial_t \mathnormal{\Pi}_{ijk} = - \varepsilon_{kmn} \partial_m \left( \varepsilon_{nrq} \mathnormal{\Pi}_{ijr} V^{(S)}_q \right) \nonumber\\
\Longrightarrow \partial_t S & = -(\mbox{curl} \ S) \times V^{(S)} + \nabla f; & \partial_t S_{ijk} = - \varepsilon_{krs} (\varepsilon_{rmn} \partial_m S_{ijn}) V^{(S)}_s + \partial_k f_{ij}\nonumber
\end{aligned}
\end{equation*}
where $V^{(S)}$ is the velocity of the g.disclination field $\mathnormal{\Pi}$, and $f$ is a free second--order tensor field to be specified that does not affect the conservation of g.disclination strength. The above kinematically fundamental evolution statement states that the $S$ field transports or `convects' only at places where a g.disclination defect is present - physically, in regions of incompatible curvature, e.g., at multi-junctions or when the boundary is curved (a straight region of a  uniform eigenwall distribution has no g.disclinations in it). If now one invokes the ansatz that the $S$ field is a gradient ($ S = \nabla \omega$), so that it \textit{cannot have any g.disclination content} and furthermore assumes that
\begin{itemize}
    \item the free energy density of the model is of the form 
    \[ 
    E(U^e) + G(Y) + \phi(S) + \epsilon^2 |\nabla S|^2
    \]
    where $E$ is the elastic energy density, $U^e$ is the elastic distortion, $G$ is an energy density related to strain gradients with $Y := \nabla U^e - S$, $\phi$ is an experimentally measured grain boundary energy density function, and $\epsilon$ a material parameter; and
    \item $U^e$ is constrained to be the field $\omega$ resulting in $\partial_Y G = 0$ (which is realizable in the absence of externally applied torques),
\end{itemize}
then the thermodynamic driving force for the field $f$ is given by 
\begin{equation}
\mbox{div}\, \partial_{\nabla \omega} \phi - 2\epsilon^2 \, \mbox{div} \, \mbox{div} \, \nabla^2 \omega.
\end{equation}
Assuming a simple one--constant mobility produces an evolution equation for $\omega$ given by
\begin{equation}\label{tensorial_form_governing_pde}
\partial_t \,\omega_{ij} = M  [ \partial_k (\partial_{\nabla \omega} \phi)_{ijk } - 2\epsilon^2 \, \partial_k \partial_k \partial_l \partial_l \, \omega_{ij}],
\end{equation}
which may be thought of as a tensorial analog of the gradient flow equations of the A--G functional. In one space dimension and time, the above equation is our \eqn{Gen_EvolutionEqn_Final} to follow. An observation worthy of note -- due to the a priori elimination of any g.disclination defects -- is that this model cannot transport or move the grain boundary /eigenwall field while allowing it to evolve `in place', and we observe this in our simulations. When the restrictive assumptions made in obtaining this `simplified' model are relaxed, the overall model naturally allows for the interaction of stress, couple stress, and dislocation plasticity with grain boundary motion, based on the fundamental kinematics of line defect dynamics. Furthermore, the operational method for defining a grain boundary in this setting by initializing the $S$ field \cite{ZHANG2018finite} shows that all 5 macroscopic parameters defining a boundary (the misorientation and the orientation of the interface normal) are naturally included in the definition of the $\phi$ function in this setting. While these are attractive features, the model (accounting for disclination defects) is `expensive' in terms of the number of physically mandated fields involved. A relatively less expensive model, still containing most of the topological complexities and capable of describing a layered `smectic' polycrystalline medium, is available in the literature \cite{ZHANG2021132828}. In this generalized sense, our work is in line with the disconnection dynamics based model of Srolovitz and co--workers \cite{wei2019continuum,HAN2021117178}, and complementary to that of a unified grain boundary evolution model with polycrystal plasticity \cite{admal2018unified}.
\section{The Continuum Grain Boundary Model}\label{sec:cgb_model}
\noindent We work in 1 space dimension ($x$) and time ($t$), with the spatial domain ($\Omega$) being the interval $[0,L]$. The free energy of the system (\textit{F}) is assumed to be of the form 

\begin{equation}\label{Overall Free Energy of the system1}
\textit{F}=\int_\Omega \varphi \left( \partial_x \theta, \partial_x^2\theta, l \right) \, dx, %%%\ \ \ \ \ \ \Omega[0,L] \hfill
\end{equation}
where $\varphi \left( \partial_x \theta, \partial_x^2\theta, l \right) $ is the overall free energy density given by
\begin{equation}\label{eq:dim_en_dens}
\varphi \left( \partial_x \theta, \partial_x^2\theta, l \right) =\phi \left(\partial_x \theta \right)+\epsilon^2 \left(\partial_x^2\theta \right)^2.
\end{equation}
Here, the length scale $l$, a typical grain boundary width, is involved in the definition of the function $\phi$ and the material parameter $\epsilon$. The latter is required on dimensional grounds and represents an energetic penalty to sharp kinks in the orientation profile.

The chosen class of the energy density is roughly justified as follows. Assuming the energy density at any point $x$ to be a function of the misorientation, $\Delta \theta (x)$, at that point over a length scale of $l$, is given by
\[
\Delta \theta (x) = \theta( x+l) - \theta(x) \approx l \partial_x \theta (x)  + \frac{1}{2} l^2 \partial^2_x \theta (x) +\cdots.
\]
Thus, for a system free energy written as $\int_\Omega \Phi (\Delta \theta) \, dx $, the energy density can be approximated to leading orders as
\[
\Phi(\Delta \theta(x)) \approx \varphi \left( \partial_x \theta(x), \partial^2_x \theta(x), l \right), 
\]
with its right hand side defined as in \eqn{eq:dim_en_dens}, and the function $\phi$ defined from experimental measurements as described in Section \ref{sec:gb_experimental_data}. Thus, our higher order regularization arises from allowing for a more accurate accounting of the effect of misorientation on system energy beyond orientation gradients.

The gradient flow evolution of the functional $F$ is expressed as
\begin{equation}\label{EvolutionEqn_General}
\partial_t \theta =-M\frac{\delta  \textit{F}}{\delta \theta},
\end{equation}
where $M$ is a scalar mobility, and it multiplies the variational derivative of the functional $F$.

The variation of $F$ in the direction $\eta$ about the state $\theta$ is expressed as
\begin{multline}\label{Variation_of_F}
\delta F = \int_{\Omega} \left[{ \partial^2_x\left({\frac{\partial \varphi}{\partial \left(\partial^2_x \theta\right)}}\right) - \partial_x\left({\frac{\partial \varphi}{\partial \left(\partial_x \theta\right)}}\right)}\right]\eta \, dx + {\eta\left[{\frac{\partial \varphi}{\partial \left(\partial_x \theta\right)} - \partial_x\left({\frac{\partial \varphi}{\partial \left(\partial^2_x \theta\right)}}\right)}\right]_{\partial \Omega}} + \\ {\partial_x \eta \left[{\frac{\partial \varphi}{\partial \left(\partial^2_x \theta\right)}}\right]_{\partial \Omega}} .
\end{multline}
The variational derivative is then given by
\begin{equation}\label{1d variation F wrt theta}
\frac{\delta F}{\delta \theta} = \partial^2_x\left({\frac{\partial \varphi}{\partial \left(\partial^2_x \theta\right)}}\right) - \partial_x\left({\frac{\partial \varphi}{\partial \left(\partial_x \theta\right)}}\right).
\end{equation}
Some of the possible boundary conditions from \eqn{Variation_of_F} are
\begin{subequations}\label{Various possible BCs}
\begin{align}
\label{particularbc-a}
\ & \theta=\textit{constant}; \partial_x \theta=0; \\
\label{particular-b}
\ & \theta=\textit{constant}; \frac{\partial \varphi}{\partial \left(\partial^2_x \theta\right)}=0 ;\\
\label{particular-c}
\ & \frac{\partial \varphi}{\partial \left(\partial_x \theta\right)} - \partial_x\left({\frac{\partial \varphi}{\partial \left(\partial^2_x \theta\right)}}\right)=0; \frac{\partial \varphi}{\partial \left(\partial^2_x \theta\right)}=0;\\
\label{particular-d}
\ &\partial_x \theta=0; \frac{\partial \varphi}{\partial \left(\partial_x \theta\right)} - \partial_x\left({\frac{\partial \varphi}{\partial \left(\partial^2_x \theta\right)}}\right)=0.
\end{align}
\end{subequations}
Substituting \eqn{1d variation F wrt theta} into \eqn{EvolutionEqn_General}, one obtains
\begin{equation}\label{Gen_EvolutionEqn_Final}
\partial_t \theta =-M\left({\partial^2_x\left({\frac{\partial \varphi}{\partial \left(\partial^2_x \theta\right)}}\right) - \partial_x J}\right),
\end{equation}
where $J$= $\frac{\partial \varphi}{\partial \left(\partial_x \theta\right)}$.
\eqn{Gen_EvolutionEqn_Final} is the governing equation for this work, along with the boundary conditions
\begin{equation}\label{BCs_1}
\begin{split}
\left(\partial_x \theta\right)_{x=0} = 0, \\
\left({\frac{\partial \varphi}{\partial \left(\partial_x \theta\right)} -\partial_x\left({\frac{\partial \varphi}{\partial \left(\partial^2_x \theta\right)}}\right) }\right)_{x=0} = 0, \\
\left(\partial_x \theta\right)_{x=L} = 0, \\
\left({\frac{\partial \varphi}{\partial \left(\partial_x \theta\right)} -\partial_x\left({\frac{\partial \varphi}{\partial\left(\partial^2_x \theta\right)}}\right) }\right)_{x=L} = 0.
\end{split}
\end{equation}
\subsection{Grain boundary energy density functions from experimental data}\label{sec:gb_experimental_data}
\noindent The grain boundary energy as a function of misorientation is a well--established experimental measurement \cite{hasson1971interfacial,miura1990determination,LI20094304}. The experimentally measured energy data of symmetrical tilt boundaries in Al is shown in \fig{hassongouxtilt}. These experimental data points can be fitted in many ways. 

In their pioneering work, Read and Shockley \cite{PhysRev.78.275} theoretically deduced a grain boundary energy density function based on the mechanics of dislocations. This model is valid for small misorientations and is consistent with a cusp in the energy density at 0 misorientation.

The variational method employed to compute interfacial energies of tilt boundaries from atomic configurations showed the existence of cusps in the $\phi$ function \cite{hasson1971interfacial}. A grain boundary energy function with cusps is proposed for fcc metals based on the essential five macroscopic degrees of freedom \cite{BULATOV2014161}. The grain boundary energy density ($\phi$) in copper evaluated by molecular dynamics (MD) simulation showed a smooth $\phi$ function \cite{takata2004grain}. Various shapes of $\phi$ are reported in the literature, and the exact shape of $\phi$ is not an established fact.

 The evolution \eqn{Gen_EvolutionEqn_Final} takes a dimensional grain boundary energy density function as input, in particular one fitted to experimentally measured grain boundary energy density data. As a representative example, the data in \fig{hassongouxtilt} corresponds to  the range of misorientations ($\Delta \theta$ = 0 to $\pi$ radians) and is converted to be a function of $(\partial_x \theta)$ as described in \eqn{misor_conv}. This function is then extended as an even function in the range \mbox{$- \pi/l$} to \mbox{$+\pi/l$}. The resulting function defines the variation in the fundamental period of a periodic function $\phi$ defined on the real line, which is used in our typical calculation. 
 
 The energy density function $\phi$ within the fundamental period is divided into $N$ segments. The $\phi(\partial_x \theta)$ is accessed by 
\begin{equation*}
    \phi(\partial_x \theta)=\mathcal{R}_q(\partial_x \theta) \ \ \ \ \ \partial_x \theta_q^s \leq \partial_x \theta \leq \partial_x \theta_q^f \ ,
\end{equation*}

where $\mathcal{R}_q(\partial_x \theta)$ is the $q^{th}$ segment of $\phi(\partial_x \theta)$ function, $\partial_x \theta_q^s$ and $\partial_x \theta_q^f$ are the start and end points of the $q^{th}$ segment, respectively, and $q$ varies from 1 to $N$.

We fit the experimental data of Hasson and Goux \cite{hasson1971interfacial} to two types of the $\phi$ functions, namely, 
\begin{itemize}
    \item a smooth energy density (SED) that we refer to as $\phi_{s}$ defined in Section \ref{sec:sed_func}, and
    \item a cusp energy density (CED), referred to as $\phi_{c}$ defined in Section \ref{sec:ced_func}.
\end{itemize} 
The SED function is a cubic spline interpolation of the entire experimental data set, capturing all energy troughs in the experimental data (assumed to be the `ground truth' here), including relatively higher energy ones. The CED function fits selected experimental data points using piecewise-concave quadratic segments, connected by smooth segments around the local minima of the energy density data. These functions are non--convex and have troughs of unequal depths. The gradient flow dynamics of these two types of the $\phi$ functions are subsequently compared.
\subsubsection{Smooth energy density (SED) function}\label{sec:sed_func}
\noindent The expression for the SED ($\phi_s$) function corresponding to the grain boundary energy density data in \fig{hassongouxtilt} is given by
\begin{equation} \label{Fitted_Poly_Grad_cspline}
 \mathcal{R}_q(\partial_x \theta) = B\left({ a_q  \left(\partial_x \theta\right)^3 + b_q \left(\partial_x \theta\right)^2 + c_q  \partial_x \theta +d_q}\right) \ \ \ \ \ \partial_x \theta_q^s \leq \partial_x \theta \leq \partial_x \theta_q^f,
\end{equation}
where $B$ is a (dimensional) material constant used to scale the data from relative to actual grain boundary energy density. It sets the energy scale of the problem. The parameters $a_q,b_q,c_q,$ and $d_q$ define the cubic polynomial for the $q^{th}$ segment of the overall spline, and ensure that the overall energy density function is a continuously differentiable function in the fundamental period.
\begin{figure}%%[H]
\centering
\includegraphics[width=1\linewidth]{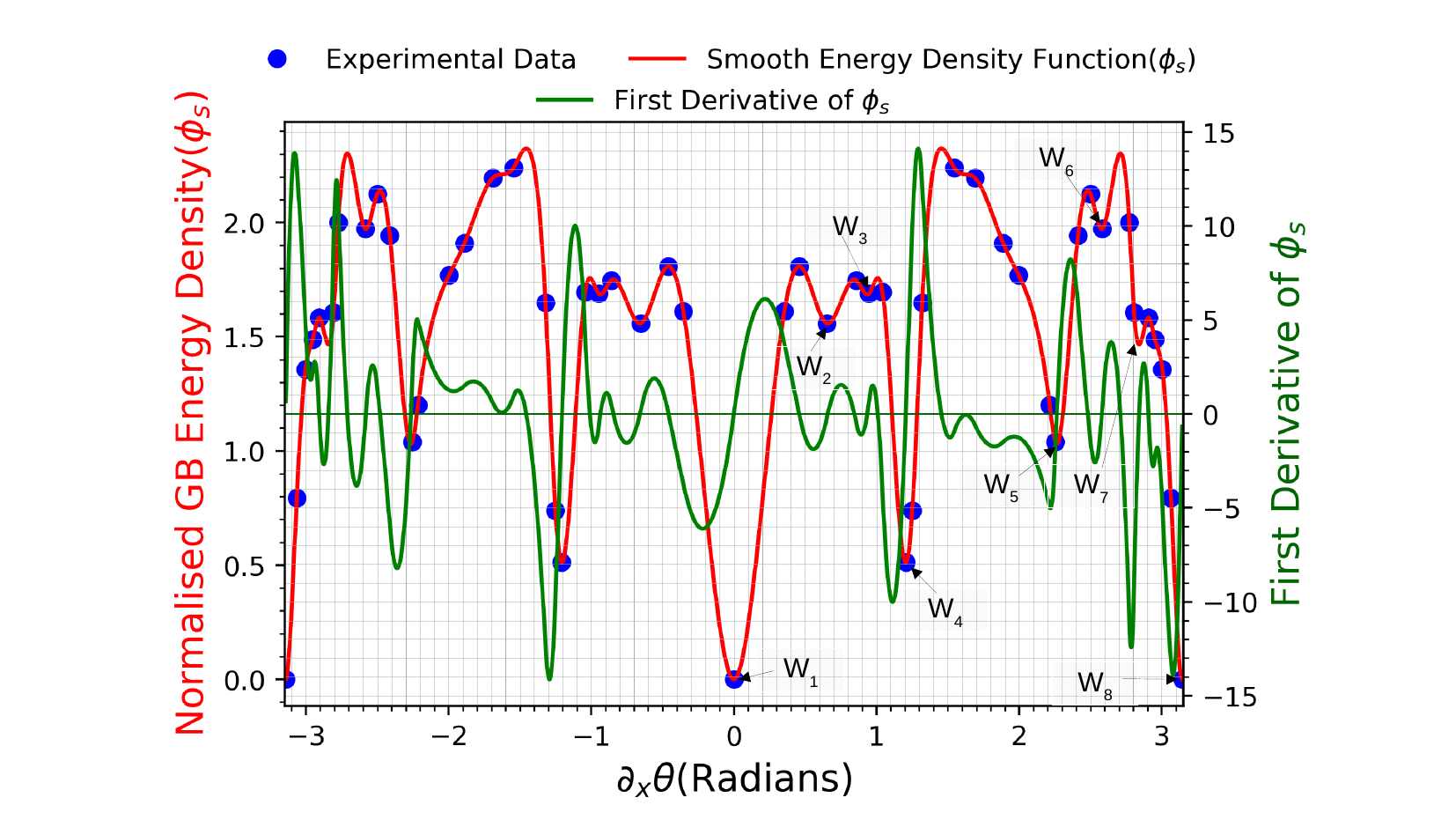}
\caption{Fitting of a smooth curve for grain boundary energy density as a function of $\partial_x \theta$ corresponding to the data obtained from experimental results \cite{hasson1971interfacial} where, $W_{1}, W_{2}, ... W_{8}$ are the various wells in the SED function.}
\label{GBEnergyVsGradTheta_Piecewise_Cubic_Splinefit}
\end{figure}
The SED function interpolating the experimental data, and its first derivative are shown in \fig{GBEnergyVsGradTheta_Piecewise_Cubic_Splinefit}. It has physical dimensions of an energy density and \eqn{Fitted_Poly_Grad_cspline} is plugged into the overall free energy of the system in \eqn{Overall Free Energy of the system1} to attain the final evolution equation corresponding to $\phi_{s}$. 
\subsubsection{Cusp energy density (CED) function}\label{sec:ced_func}
\noindent Experimentally measured data is often fitted to (logarithmic) cusp type grain boundary energy density functions in the literature \cite{hasson1971interfacial,miura1990determination}, following the Read--Shockley model \cite{PhysRev.78.275}. 

As already mentioned, the CED function consists of piecewise-concave quadratic segments, connected by smooth segments around local minima of the energy density data. Let $A$ be the set of indices that identify the $5^{th}$ degree polynomials around the local minima, and let $C$ be the set of indices that identify the piecewise-concave quadratic segments, with $A$ and $C$ being disjoint sets. Then the expression for the CED  function, $\phi_c$, is
\begin{equation}\label{GB ED CGB Cusp Model_1}
\frac{1}{B} \mathcal{R}_q(\partial_x \theta)=
\begin{cases}
    a_q  \left(\partial_x \theta\right)^5 + b_q \left(\partial_x \theta\right)^4 + c_q  \left(\partial_x \theta\right)^3 \\
    + d_q \left(\partial_x \theta\right)^2 + e_q \partial_x \theta + f_q, \qquad q \in A \\
    a_q  \left(\partial_x \theta\right)^2 + b_q \partial_x \theta + c_q,\qquad q \in C 
\end{cases} \qquad \partial_x \theta_q^s \leq \partial_x \theta \leq \partial_x \theta_q^f.
\end{equation}

\begin{figure}%%[h]
\centering
\includegraphics[width=1.0\linewidth]{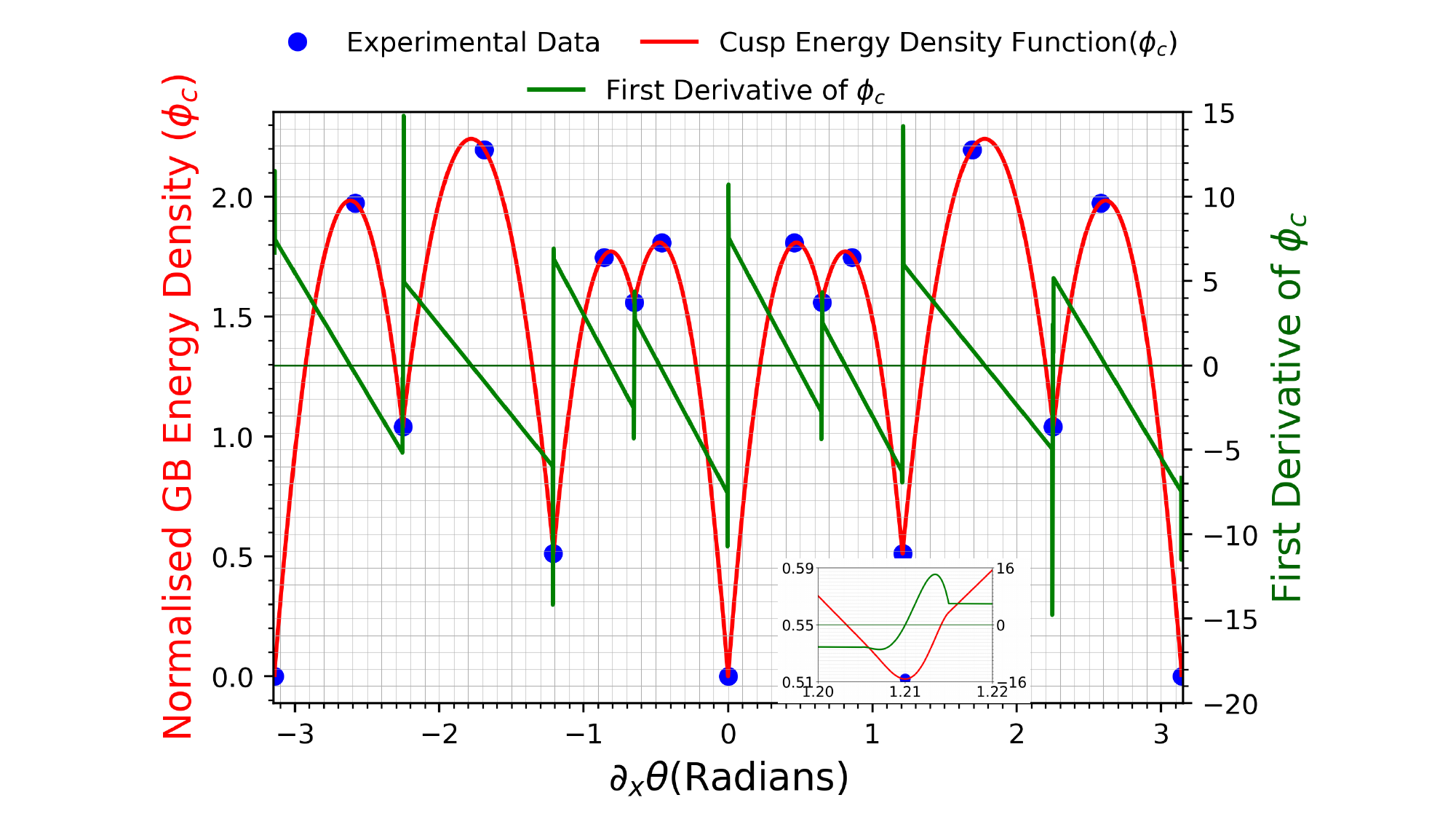}
\caption{Cusp type fit for grain boundary energy density as a function of \mbox{$\partial_x \theta$} corresponding to the data obtained from experimental results \mbox{\cite{hasson1971interfacial}}. The subplot inside the figure shows the smooth transition of \mbox{$\phi'_c$} near the \mbox{$\partial_x \theta=1.2$} trough.}
\label{Fig4}
\end{figure}

The function $\phi_c$ consists of $N=2p-1$ segments where $p$ is the total number of troughs. \fig{Fig4} shows the concave quadratic polynomials, each between two troughs of the experimentally measured grain boundary energy density data of \cite{hasson1971interfacial}. Each concave quadratic polynomial passes through three experimental data points and the corresponding coefficients of the concave quadratic polynomials are $a_q,b_q$ and $c_q$, $q \in C$. 
Each $5^{th}$ degree polynomial is defined between two concave quadratic polynomials in a small range of $\partial_{x} \theta=1\times 10^{-2}$ around an energy density trough.  Their values and slopes are continuous with the adjacent concave quadratic polynomials at the junctions. Furthermore, the fitted $\phi_c$ is designed to align with experimental data at the trough, and its slope is set to zero at this point.  These six constraints collectively necessitate the use of at least a fifth-degree polynomial. This rigorous selection criterion underscores the rationale for employing a fifth-degree polynomial between two concave quadratic polynomials. These $5^{th}$ degree polynomials with coefficients $a_q,b_q,c_q,d_q,e_q$ and $f_q$, $q \in A$, also pass through the experimentally measured data points at troughs, with 0 slope there.

The subplot in \fig{Fig4} shows the continuous and smooth transition of $\phi'_c$ (derivative) at the $\partial_{x} \theta=1.2$ trough. As with the SED function, the CED function expressed in \eqn{GB ED CGB Cusp Model_1} is plugged into the overall free energy of the system in \eqn{Overall Free Energy of the system1} to attain the final evolution equation corresponding to $\phi_{c}$.

The governing equation shown in \eqn{Gen_EvolutionEqn_Final} can be non--dimensionalized as
\begin{equation}\label{EvolutionEqn_Final_cspline}
\partial_{\widetilde{t}} \theta =-\alpha \partial^4_{\widetilde{x}} \theta + \partial_{\widetilde{x}} {\widetilde{J}}\left(\partial_{\widetilde{x}} \theta\right),
\end{equation}
where, 
\begin{equation*}
x=l\widetilde{x}, \qquad t=\frac{\widetilde{t}}{MB}, \qquad \partial_x \theta = \frac{1}{l} \partial_{\widetilde{x}}\theta, \qquad \partial_t \theta = MB \partial_{\widetilde{t}} \theta, \qquad \alpha=\frac{2 \epsilon^2}{B l^4}.
\end{equation*}
where we recall that $l$ represents the width of the grain boundary, $M$ is the grain boundary mobility, $\epsilon$ is the strength of the energy penalty to sharp kinks in the orientation profile. The $\ \widetilde{}$ over a variable represents the non--dimensionalized quantity of the corresponding variable. 

\textit{Henceforth, we work solely with the non-dimensional problem, but the overhead $\ \widetilde{}$ are dropped for notational convenience.}

In case the $\theta$ profile has enough smoothness, it is possible to operate on the $J$ term with the gradient in \eqn{EvolutionEqn_Final_cspline} using the product rule. This produces the form
\begin{equation}\label{EvolutionEqnJ2_2}
\partial_{t} \theta =-\alpha \partial^4_{x} \theta + J_2\left(\partial_{x} \theta\right)  \partial^2_{x} \theta.
\end{equation}
where $J_2 (\partial_{x} \theta) := \left(\frac{\partial^2}{\partial (\partial_{x} \theta)^2} \phi \right)(\partial_{x} \theta)$. There are natural conditions where the $\theta$ profile could have kinks in it, and there is not enough smoothness in the $\theta$ profile to use the product rule. In such situations, the equation in the form \eqn{EvolutionEqnJ2_2} does not make sense and its use produces spurious results as analyzed in \ref{appendix:b}.

Therefore, the evolution equation \eqn{EvolutionEqn_Final_cspline} is used in the present work with a corresponding numerical implementation that can handle the kinks and preserves the jump conditions, as demonstrated in Section \ref{section:results_discussion}.

\section{Numerical Implementation}\label{section:numericalimplementation}
\noindent The \eqn{EvolutionEqn_Final_cspline} is numerically discretized using the central difference formula in space and by the forward Euler method in time and the corresponding fully explicit numerical discretization is given by
\begin{equation}\label{Numerical_Discretization sform}
\frac{\theta_i ^{t+dt}-\theta_i ^{t}}{dt} = -\alpha \left[ {\frac{\theta_{i+2} ^t-4\theta_{i+1} ^t+6\theta_{i} ^t-4\theta_{i-1} ^t+\theta_{i-2} ^t}{(\Delta x)^4}}\right] + \frac{J\left({\partial_{x} \theta_{i+\frac{1}{2}}^t}\right)-J\left({\partial_{x} \theta_{i-\frac{1}{2}}^t}\right)}{\Delta x},
\end{equation}
where,
\begin{equation*}
\partial_{x} \theta_{i+\frac{1}{2}}^t = (\partial_{x} \theta)_{i+\frac{1}{2}}^t = \left({\frac{\theta_{i+1}^t-\theta_{i}^t}{\Delta x}}\right),
\end{equation*}
\begin{equation*}
\partial_{x} \theta_{i-\frac{1}{2}}^t = (\partial_{x} \theta)_{i-\frac{1}{2}}^t = \left({\frac{\theta_{i}^t-\theta_{i-1}^t}{\Delta x}}\right),
\end{equation*}
\renewcommand{\thefootnote}{\fnsymbol{footnote}} $\Delta x$ is the uniform spatial grid size, $i$ is the node\footnote[2]{Nodes are the discrete points located at $ x = i \Delta x$ where the discrete field $\theta$ is determined.} number, and $dt$ is the time step. $t$ and $t+dt$ represent discrete time instants. $\theta_i ^{p}$ represents the value of $\theta$ at the $i^{th}$ node at time $p$. The rest of the terms in \mbox{\eqn{Numerical_Discretization sform}} follow a similar analogy. 

The fourth--order term ($\partial^4_{x} \theta$) is evaluated at the nodes whereas the $J$ term is evaluated on the $\partial_{x} \theta$ calculated at the cell centers\footnote[3]{The region between two nodes is considered a cell. The cell center is the midpoint of a cell.}$\left(\partial_{x} \theta\right)_{i\pm \frac{1}{2}}$ to avoid the problem of numerical instability due to the kinks. \mbox{\eqn{Numerical_Discretization sform}} is rewritten and used in the form
\begin{multline}\label{Numerical_Discretization2 sform}
\theta_i ^{t+dt}=\theta_i ^t \left[{1-\frac{6\alpha dt}{(\Delta x)^4}}\right] + \theta_{i+1} ^t \left[{\frac{4\alpha dt}{(\Delta x)^4}}\right] + \theta_{i+2} ^t \left[{\frac{-\alpha dt}{(\Delta x)^4}}\right] + \theta_{i-2} ^t \left[{\frac{-\alpha dt}{(\Delta x)^4}}\right] \\
 + \theta_{i-1} ^t \left[{\frac{4\alpha dt}{(\Delta x)^4}}\right] +\frac{dt}{\Delta x}\left(J\left(\partial_{x} \theta_{i+\frac{1}{2}}^t\right)-J\left(\partial_{x} \theta_{i-\frac{1}{2}}^t\right)\right).
\end{multline}
Forward and backward difference formulae are used at the boundaries. 

We note that for $\alpha = 0$,  \eqn{Numerical_Discretization2 sform} imposes the jump condition $J|_{x^+} = J|_{x^-}$, where $J|_{x^\pm}$ are the limits of the evaluation of $J$ from the right and left of $x$, respectively, for \eqn{EvolutionEqn_Final_cspline} (assuming $\theta_t$ to be an intergrable function). This is significant when there is a discontinuity in the function $x \mapsto J(x)$ at the point $x$, as happens when there is kink in $x \mapsto \theta(x)$ at $x$. In these situations, our code imposes this jump condition exactly for a kink at node $i$. Of course, when $\alpha \neq 0$ the term remains as a component of a consistent discretization of the PDE we wish to solve to explore our model.
\subsection{Von Neumann stability analysis}
\noindent Von Neumann stability analysis provides time-stepping constraints for discretized constant coefficient linear partial differential equations, including finite difference schemes \cite{isaacson1994analysis}. The $J$ term in \eqn{EvolutionEqn_Final_cspline} is strongly nonlinear. Hence, the strategy we adopt is to  formally write our equation in the form
\begin{equation}\label{EvolutionEqnJ2}
\partial_{t} \theta =-\alpha \partial^4_{x} \theta + J_2\left(\partial_{x} \theta\right)  \partial^2_{x} \theta,
\end{equation}  
presume it to be a constant coefficient, perform a von Neumann stability analysis, deduce the time step constraint(s) corresponding to the $J_2$ value at each node, and accept the minimum of these estimates over the domain. Here, $J_2$ at node $i$ is defined as $J_2( (\partial_x \theta)_i)$ where 
    $(\partial_{x} \theta)_i = (\theta_{i+1}-\theta_{i-1})/(2 \Delta x)$.
The growth factor for the explicit scheme \mbox{\eqn{Numerical_Discretization sform}} is obtained from the expression

\begin{equation*}
\theta_k ^{t+dt}=   \left({{1-\frac{\alpha dt (2\cos(2k\Delta x) -8\cos(k\Delta x)+6)}{(\Delta x)^4}+\frac{J_2 dt (2\cos(k\Delta x) -2)}{(\Delta x)^2}}}\right) \theta_k ^t,
\end{equation*}
resulting in the stability constraint given by

\begin{equation}\label{Stability_Condition}
\left |{{1-\frac{\alpha dt (2\cos(2k\Delta x) -8\cos(k\Delta x)+6)}{(\Delta x)^4}+\frac{J_2 dt (2\cos(k\Delta x) -2)}{(\Delta x)^2}}} \right| \leq 1 + \mathcal{O}(dt) \quad \forall \ 0 \leq k \leq \infty
\end{equation}
(well-posed growth of the solution has to be allowed for $J_2 < 0, \alpha > 0$) where,
\begin{equation*}
0 \leq (2\cos(2k\Delta x) -8\cos(k\Delta x)+6) \leq 16; \qquad -4 \leq (2\cos(k\Delta x) -2) \leq 0.
\end{equation*}
The stability constraints for the scheme are given by ($\Delta x > 0$): 
\begin{equation}\label{Overall_Stability_Constraints_1}
dt\leq
\begin{cases}
    \frac{\Delta x^2}{2J_2},& \alpha=0, \ J_2\geq0\\
    \text{No viable time step},& \alpha=0, \ J_2<0\\
    \frac{\Delta x^4}{8\alpha},& \alpha>0, \ J_2=0\\
    \frac{\Delta x^4}{8\alpha},& \alpha>0, \  0 < \Delta x \leq \sqrt{\frac{4\alpha}{|J_2|}}, \ J_2<0\\
    \frac{\Delta x^4}{2(4\alpha+|J_2|\Delta x^2)},& \alpha>0, \ \ J_2\geq0.\\
\end{cases}
\end{equation}

As already mentioned, the time step ($dt$) is calculated at all the nodes according to the stability constraints in \eqn{Overall_Stability_Constraints_1}, and the least value among all of them is used as the stable time step to solve \eqn{Numerical_Discretization2 sform}.
\section{Results and Discussion}\label{section:results_discussion}

\noindent The results presented in this section are generated using the explicit numerical scheme described in Section \ref{section:numericalimplementation}. The adaptive time stepping ($dt$) expressed in \eqn{Overall_Stability_Constraints_1} is multiplied with a factor of 0.75 for all simulations. The simulations consistently utilize a uniform mesh size ($\Delta x$) of 0.1, unless explicitly stated otherwise.

Before discussing detailed simulations, we briefly outline the goals of the exercise we undertake in this Section. 

First, we note that the two energy functions chosen, i.e., $\phi_c$ and $\phi_s$, are fitted to a \textit{single} experimental data set. In particular, both energy density functions have an identical set of local minima. With this in mind, a primary question we explore is to what extent the `nonequilibrium shape' of the energy density functions, i.e., all characteristics of the function beyond the specification of the local minima, affect the prediction of equilibria and time-dependent approach to it.

Second, given the nature of the energy functions employed, grain boundary equilibria can at most be metastable states of the simulation systems considered. We computationally check the existence and stability of such metastable states.

Various evolutions and their equilibria, the latter always interpreted `loosely' as maximum changes in the orientation in the whole domain being less than a specified (non--dimensional) threshold that is deemed physically reasonable, are discussed. These simulations recover idealized features of real physical systems as follows:
\begin{itemize}
    \item equilibrium high--angle grain boundaries (HAGBs) observed in actual polycrystalline materials;
    \item grain rotation observed in real polycrystalline materials;
    \item grain growth during annealing;
    \item heavily deformed (static) microstructure often observed after the deformation process in actual polycrystalline materials;
    \item dislocation walls and polygonized domains in equilibria.
\end{itemize}

\subsection{Initial condition with a single transition layer in a deep energy density  well}\label{IC_single_TL}
\noindent Polycrystalline materials are found often in states that have grain boundaries between grains that appear to be static. This case study aims to test the model in producing an equilibrium transition layer between two regions of piecewise constant orientation, i.e. a bicrystal. According to the literature, a misorientation of $\Delta \theta$ $\leq$ $10^{\circ}$ is considered as the grain interior (including the sub-grain boundaries) and $\Delta \theta > 10^{\circ}$ is considered as the grain boundary (HAGBs) \cite{HUMPHREYS200491}. These $\Delta \theta$ ranges are converted to $\partial_x \theta$ as described in \eqn{misor_conv} and $\partial_x \theta$ is non-dimensionalised as shown in \eqn{EvolutionEqn_Final_cspline}. Now, the grain interior is defined as $\partial_x \theta \leq 0.1745$ (analogous  to $\Delta \theta$ $\leq10^{\circ}$) and a grain boundary is defined as $\partial_x \theta > 0.1745$ (analogous to $\Delta \theta$ $>10^{\circ}$).

\begin{figure}%%[H]
    \centering
    \begin{subfigure}{0.475\textwidth}
        \centering
        \includegraphics[width=1\linewidth]{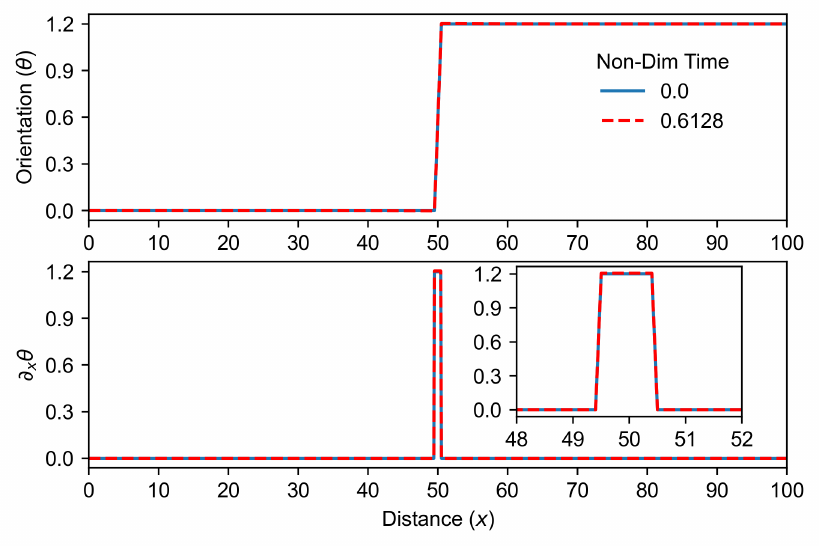}
        \caption{}
        \label{Fig5a}
    \end{subfigure}
    \hfill
    \begin{subfigure}{0.49\textwidth}
        \centering
        \includegraphics[width=1\linewidth]{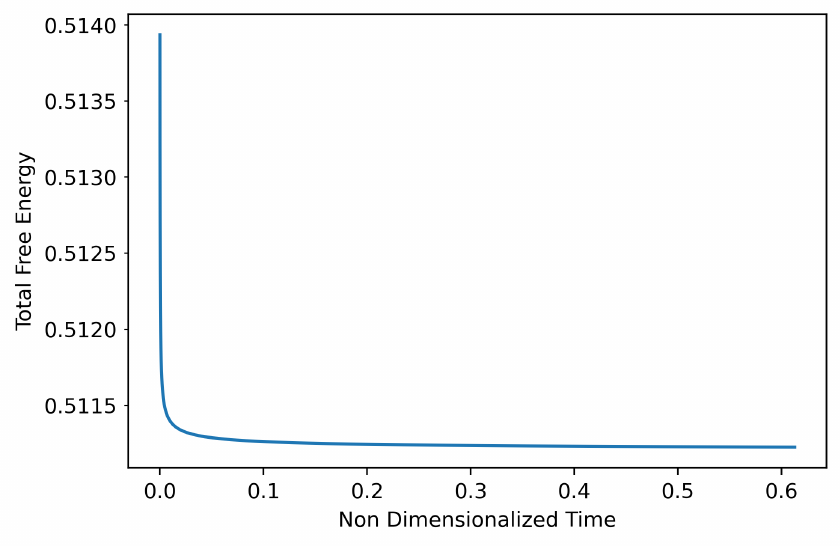}
        \caption{}
        \label{EnergyVsTime_Explicit_Cspline_J1from_gradtheta_1p2_alpha_0_TT_0p6128}
    \end{subfigure}
    \hfill
    \begin{subfigure}{0.49\textwidth}
        \centering
        \includegraphics[width=1\linewidth]{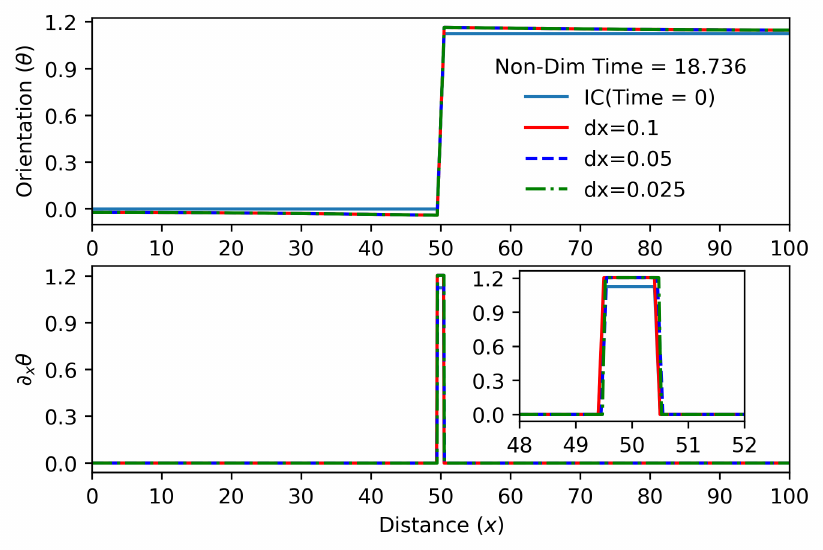}
        \caption{}
        \label{Fig5c}
    \end{subfigure}
    \hfill    
    \begin{subfigure}{0.49\textwidth}
        \centering
        \includegraphics[width=1\linewidth]{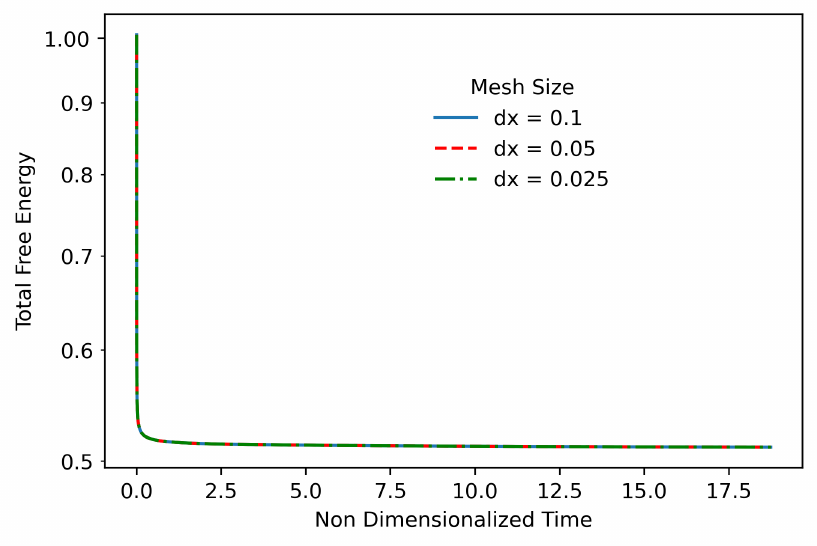}
        \caption{}
        \label{Fig5d}
    \end{subfigure}
    \hfill
    \begin{subfigure}{0.49\textwidth}
        \centering
        \includegraphics[width=1\linewidth]{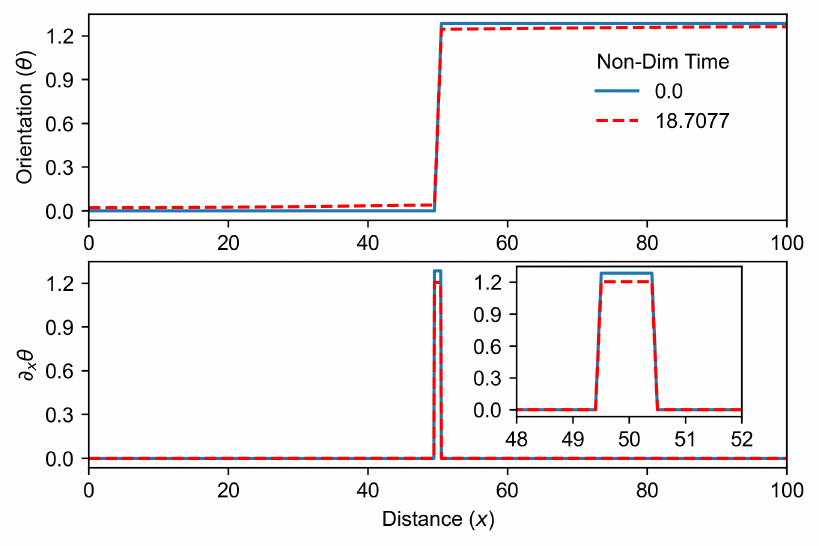}
        \caption{}
        \label{Fig5e}
    \end{subfigure}
    \hfill
    \begin{subfigure}{0.49\textwidth}
        \centering
        \includegraphics[width=1\linewidth]{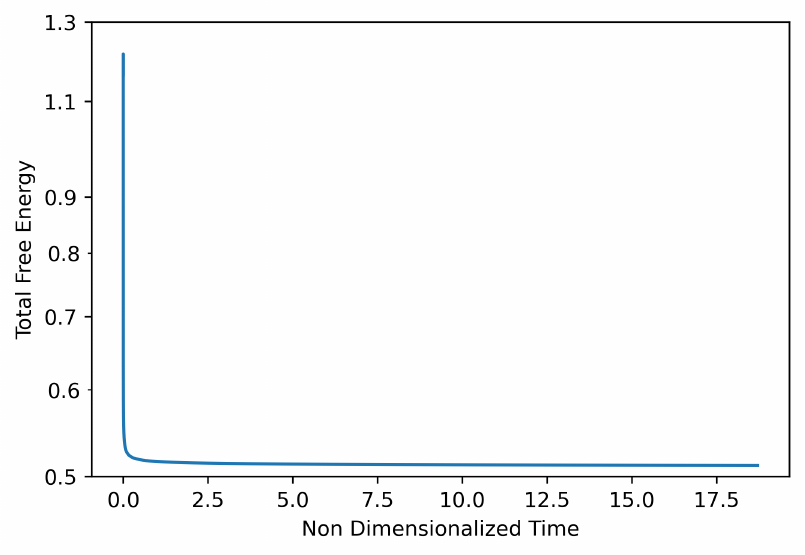}
        \caption{}
        \label{EnergyVsTime_cspline_explicit_J1_gt_1p285_alpha_0_TT_18p707}
    \end{subfigure}
    \caption{The left column of the figures shows the temporal evolution of $\theta$ for (a) $\partial_x \theta$=1.2, (c) $\partial_x \theta$=1.125 and (e) $\partial_x \theta$=1.285 narrow transition layer ($l=1$) initial conditions and the evolution of corresponding total free energies are shown in the right column of figures in (b) (d) and (f), respectively. The energy density function is SED and $\alpha=0$. The system produces the final equilibrium profile corresponding to $max|(\theta^{t+dt}-\theta^{t})/dt| < 1\times 10^{-3}$ and $|(F^{t+dt}-F^t)/dt| < 5\times 10^{-5}$.}
    \label{single_TL_1p2_1p125_1p285_IC}
\end{figure}

With this notion, the initial condition shown in \fig{Fig5a} represents a bicrystal configuration with a transition layer/grain boundary in the center where $\partial_x \theta=1.2$, corresponding to a local energy minimum of the SED energy shown in \fig{GBEnergyVsGradTheta_Piecewise_Cubic_Splinefit}. All the results in \fig{single_TL_1p2_1p125_1p285_IC} are generated using this energy function. 

We note that the initial condition contains a genuine discontinuity in orientation gradient at the kinks, and both grains and the grain boundary interior belong to convex parts of the energy density function (i.e., take orientation values belonging to convex regions of the energy density function). 

The system is allowed to evolve without the influence of the higher--order smoothing term in \eqn{EvolutionEqn_Final_cspline} by choosing $\alpha=0$. From \fig{Fig5a}, it is evident that the orientations do not change significantly as a function of time and achieve the prescribed equilibrium criterion defined by $max_x| (\theta^{t+dt}(x)-\theta^{t}(x))/dt|<1\times 10^{-3}$, at $t = 0.61$.

It is also demonstrated from \fig{EnergyVsTime_Explicit_Cspline_J1from_gradtheta_1p2_alpha_0_TT_0p6128} that the difference in total energy between the initial and the final configuration is not significant. It indicates that the chosen initial condition is close to the equilibrium configuration. From \fig{EnergyVsTime_Explicit_Cspline_J1from_gradtheta_1p2_alpha_0_TT_0p6128}, also note that the energy decrease is initially steep and the rate of change decreases and tends to become flat at $t \approx 0.61$. It indicates that there is a negligible driving force for the system to evolve further and equilibrium is reached.

An identical study is performed by fixing the equilibrium criteria as $max_x| (\theta^{t+dt}(x)-\theta^{t}(x))/dt| < 1\times 10^{-14}$ and energy rate $|(F^{t+dt}-F^t)/dt| < 1 \times 10^{-14}$ where $F^{t}$ and $F^{t+dt}$ are the total free energies of the system at the previous and current time step, respectively. The results replicate the earlier results obtained with the equilibrium criterion $max_x|(\theta^{t+dt}(x)-\theta^{t}(x))/dt|<1\times 10^{-3}$. Therefore, the equilibrium criterion $max|(\theta^{t+dt}-\theta^{t})/dt|<1\times 10^{-3}$ is used for the rest of the studies unless stated explicitly (and we use the shorthand $max$ to indicate $max_x$).
\begin{figure}%%[H]
    \centering
    \begin{subfigure}{0.49\textwidth}
        \centering
        \includegraphics[width=1\linewidth]{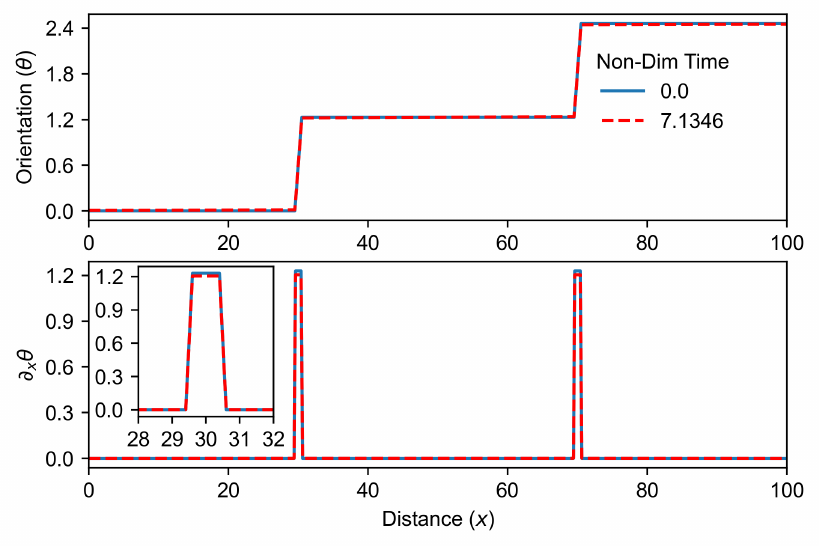}
        \caption{}
        \label{Fig6a}
    \end{subfigure}
    \hfill
    \begin{subfigure}{0.475\textwidth}
        \centering
        \includegraphics[width=1\linewidth]{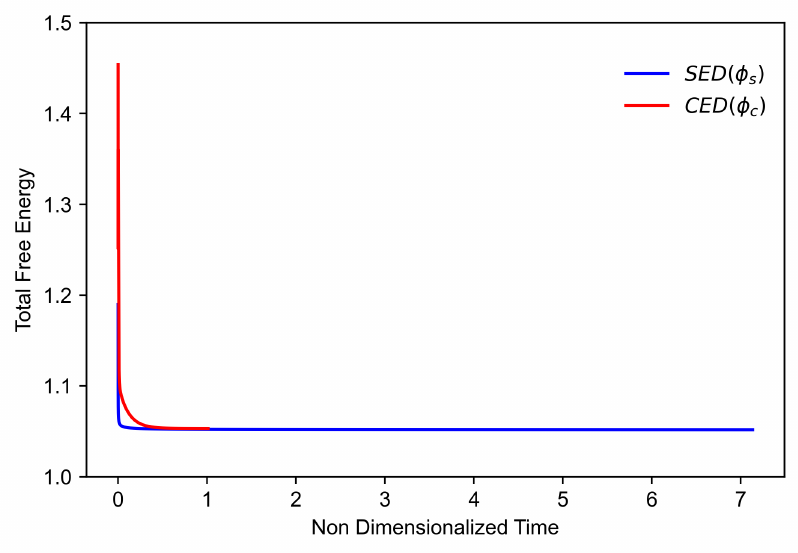}
        \caption{}
        \label{Fig6b}
    \end{subfigure}
    \caption{Temporal evolution of \mbox{$\theta$} is shown in (a) for \mbox{$\partial_x \theta=1.23$} narrow transition layer (\mbox{$l=1$}) with \mbox{$\alpha=1\times 10^{-3}$}. The total free energy evolution is shown in (b). The system produces the final equilibrium profile corresponding to \mbox{$max|(\theta^{t+dt}-\theta^{t})/dt|<1\times 10^{-3}$} and \mbox{$|(F^{t+dt}-F^t)/dt| < 5.1 \times 10^{-5}$}.}
    \label{Multiple_GB_IC_gt_1p23_0p001_SED_Expli_J1}
\end{figure}
The simulations in \fig{Fig5c} and \fig{Fig5e} aim to simulate idealized grain rotation in 1--D that is reminiscent of grain rotation observed in real polycrystalline materials \cite{HARRIS19982623,margulies2001situ}. \fig{Fig5c} shows the results corresponding to $\alpha=0$ and an initial condition with $\partial_x \theta=1.125$ at the transition layer. This is slightly away from the $\partial_x \theta=1.2$ equilibrium trough but still in the convex region of the SED function. The grains (flat regions in $\theta$) start moving away from each other by rotation. The misorientation between grains increases due to rotations of grains and produces an equilibrium $\partial_x \theta$ = 1.2 transition layer at $t = 18.73$. Similarly, for the initial condition of $\partial_x \theta=1.285$ at the transition layer (see \fig{Fig5e}), the two grains rotate towards each other during evolution and equilibrium is achieved at $t = 18.7$ (see \fig{EnergyVsTime_cspline_explicit_J1_gt_1p285_alpha_0_TT_18p707}). The final $\partial_x \theta$ at the transition layer is 1.2 which is the same as in the other two cases. The impact of mesh size on grain rotation is investigated by simulating the initial condition shown in \mbox{\fig{Fig5c}} with different mesh sizes. The results indicate that mesh size does not affect grain rotation. The total energy plots shown in \fig{Fig5d} and \fig{EnergyVsTime_cspline_explicit_J1_gt_1p285_alpha_0_TT_18p707} follow the same trend as in \fig{EnergyVsTime_Explicit_Cspline_J1from_gradtheta_1p2_alpha_0_TT_0p6128}. The values of the total energy converge asymptotically to \mbox{$\approx 0.52$} and align with the grain boundary energy corresponding to \mbox{$\partial_x \theta \approx$ 1.2} (\mbox{$W_4$}) well in \mbox{\fig{GBEnergyVsGradTheta_Piecewise_Cubic_Splinefit}},  when the system achieves equilibrium in all the three $\alpha=0$ cases. Analogous to the total free energy of single grain boundary in \mbox{\fig{single_TL_1p2_1p125_1p285_IC}}, \mbox{\fig{Multiple_GB_IC_gt_1p23_0p001_SED_Expli_J1}} depicts an initial condition with two grain boundaries positioned far apart. As these boundaries are non-interacting and distant, their corresponding total energy asymptotically converges to $\approx 1.05$ in \mbox{\fig{Multiple_GB_IC_gt_1p23_0p001_SED_Expli_J1}} for both SED and CED functions, mirroring the total free energy of two individual grain boundaries in \mbox{\fig{single_TL_1p2_1p125_1p285_IC}}.

\begin{figure}%%[H]
    \centering
    \begin{subfigure}{0.49\textwidth}
        \centering
        \includegraphics[width=1\linewidth]{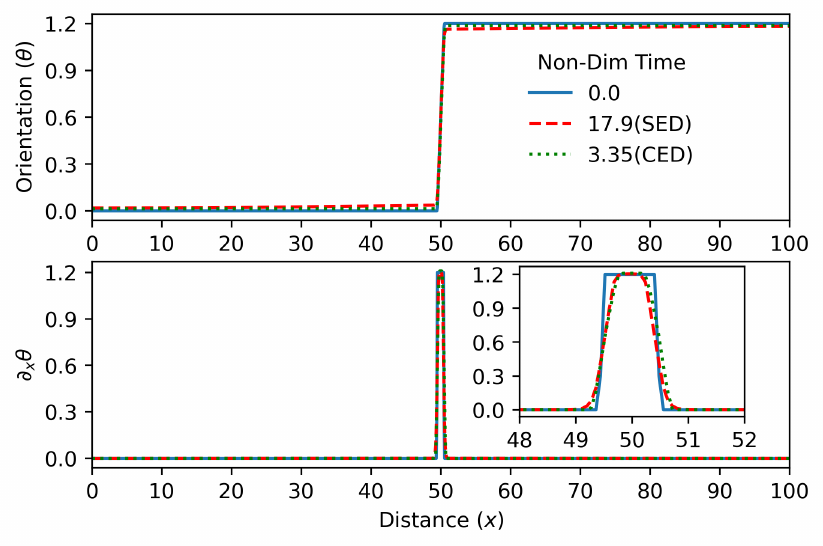}
        \caption{}
        \label{Fig7a}
    \end{subfigure}
    \hfill
    \begin{subfigure}{0.475\textwidth}
        \centering
        \includegraphics[width=1\linewidth]{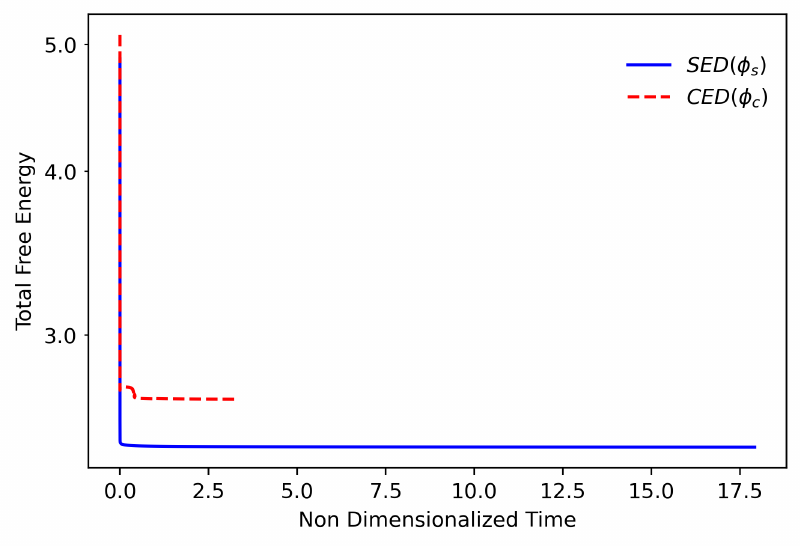}
        \caption{}
        \label{EnergyVsTime_Cspline_Cusp_Combined_gt_1p2_a2_0p37_Explicit}
    \end{subfigure}
    \caption{Temporal evolution of \mbox{$\theta$} is shown in (a) for \mbox{$\partial_x \theta=1.2$} narrow transition layer (\mbox{$l=1$}) with \mbox{$\alpha=0.37$} and a mesh size ($dx$) of 0.08. The total free energy evolution is shown in (b). The system produces the final equilibrium profile corresponding to \mbox{$max|(\theta^{t+dt}-\theta^{t})/dt|<1\times 10^{-3}$} and \mbox{$|(F^{t+dt}-F^t)/dt| < 3.12 \times 10^{-4}$}.}
\end{figure}

To analyze the effect of $\alpha > 0$ on the evolution of the orientation field, we experiment with various $\alpha$ values, ranging from 0 to 5, to identify an optimal $\alpha$ value that would render $\partial_x J$ more influential than $\partial_x^4 \theta$ in \mbox{\eqn{EvolutionEqn_Final_cspline}}. This is done because $\partial_x J$ is considered to be the primary factor in grain boundary evolution based on experimental data, while the inclusion of the fourth-order term is necessary to address stability concerns related to kinks. Thus, our objective is to strike a balance by determining an $\alpha$ value that effectively incorporates both the experimental data and stability concerns. The $\alpha = 0.37$ is one of the values that serve this purpose. Therefore,  $\alpha = 0.37$ is considered and the initial condition shown in \fig{Fig5a} is used. The subplot in \fig{Fig7a} shows that the transition of $\partial_x \theta$ is smooth from 0 to 1.2 whereas, for $\alpha=0$, the transition of $\partial_x \theta$ is sharp from 0 to 1.2 as shown in \fig{Fig5a}. These results demonstrate that $\alpha=0.37$ smooths the $\theta$ profile by eliminating kinks that is analogous to grain boundary relaxation. 

The initial condition described in \fig{Fig7a} is also used to study the evolution of $\partial_x \theta=1.2$ transition layer with $\alpha=0.37$ corresponding to the CED function. The equilibrium transition layers shown in \fig{Fig7a} are similar but the time to reach equilibrium, when using the CED function, is $(1/5)^{th}$ of that in the case when SED function is used. The reason for this in all likelihood is that the CED function shown in \fig{Fig4} has higher non--convexity than the SED function shown in \fig{GBEnergyVsGradTheta_Piecewise_Cubic_Splinefit}. \fig{EnergyVsTime_Cspline_Cusp_Combined_gt_1p2_a2_0p37_Explicit} shows the total energy evolution corresponding to \fig{Fig7a}. The equilibrium energy for the CED function is higher than that for the SED function. These results signify that the shape of the chosen energy density function influences the dynamics of the evolution to reach the equilibrium profile as well as the final equilibrium energy of the system.

The initial condition presented in \mbox{\fig{Fig7a}} is also utilized to examine the evolution of the \mbox{$\partial_x \theta=1.2$} transition layer with \mbox{$\alpha=0.37$} and boundary condition described in \mbox{\eqn{particularbc-a}} corresponding to the SED function. The grain boundary evolution closely resembles the final \mbox{$\theta$} configuration in \mbox{\fig{Fig7a}}, and within the grains, a slight gradient is observed to accommodate the applied boundary condition described in \mbox{\eqn{particularbc-a}}. Additionally, we employed the final equilibrium $\theta$ profile obtained in \mbox{\fig{Fig7a}} as the initial condition with the boundary condition in \mbox{\eqn{particularbc-a}} to verify whether the model attains the equilibrium configuration depicted in \mbox{\fig{Fig7a}} under the boundary condition in \mbox{\eqn{particularbc-a}}.  The initial and final conditions overlap, indicating minimal change. The total free energy remains nearly constant, affirming that the metastable state in \mbox{\fig{Fig7a}} can be achieved with the boundary condition in \mbox{\eqn{particularbc-a}} as well.

From these results, it can be concluded that the present model with the SED and the CED functions is capable of simulating rotation in 1--D between grains by reducing the overall energy of the system.

Finally, we show that, with $\alpha > 0$, this grain boundary/transition layer is a stable configuration to small amplitude, high wave-number perturbations. This is demonstrated in \ref{appendix:a}.
\subsection{Initial condition with multiple transition layers at troughs}
\noindent Polycrystalline materials subjected to some form of activation energy such as temperature and deformation lead to grain growth by the combined effect of grain boundary motion and grain rotation to eliminate the orientation difference between them. These orientation differences cause the formation of grain boundaries leading to excess energy in the polycrystalline system. The grain growth occurs to reduce the overall energy of the system by either decreasing the grain boundary area or eliminating the grain boundaries \cite{HUMPHREYS200491}. This study aims to show the capability of the model to simulate grain growth by grain rotation in 1--D.

\begin{figure}%%[H]
    \centering
    \begin{subfigure}{0.49\textwidth}
        \centering
        \includegraphics[width=1\linewidth]{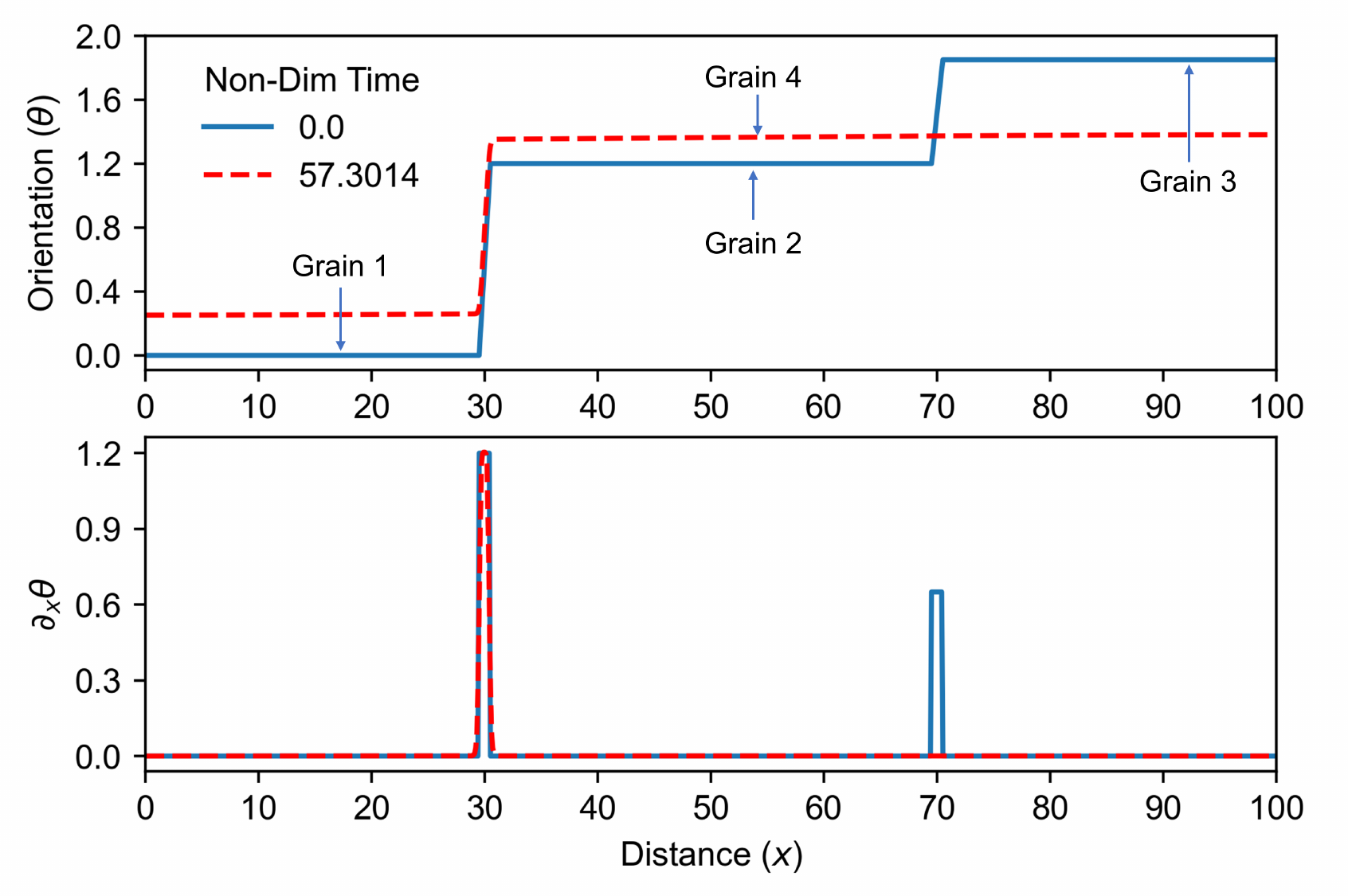}
        \caption{}
        \label{Fig8a}
    \end{subfigure}
    \hfill
    \begin{subfigure}{0.475\textwidth}
        \centering
        \includegraphics[width=1\linewidth]{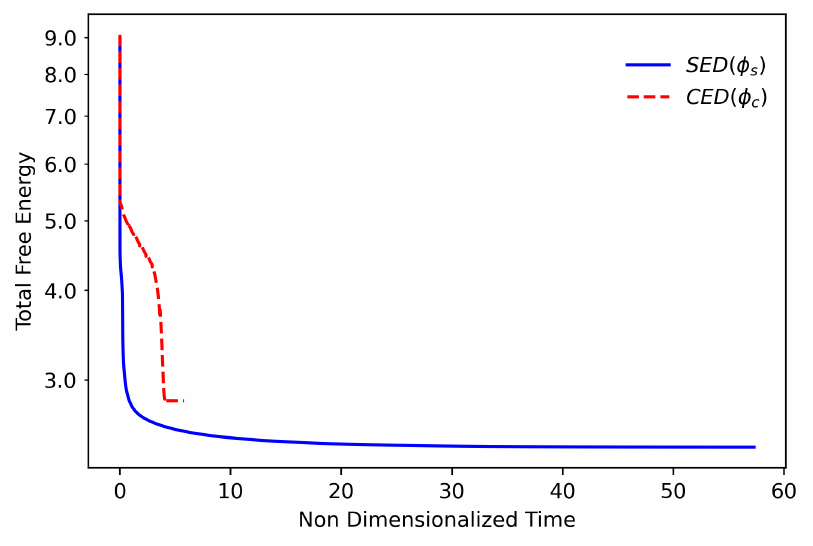}
        \caption{}
    \end{subfigure}
    \caption{Temporal evolution of $\theta$ is shown in (a) having multiple transition layers ($l=1$) using the SED function and $\alpha=0.37$. The evolution of total free energy is shown in (b). The system produces the final equilibrium profile corresponding to $max|(\theta^{t+dt}-\theta^{t})/dt|<1\times 10^{-3}$ and $|(F^{t+dt}-F^t)/dt| < 5 \times 10^{-5}$.}
\end{figure}

The initial condition in \fig{Fig8a} shows 3 grains with different orientations where the $\partial_x \theta$ between grain 1 and grain 2 is 1.2 and between grain 2 and grain 3 it is 0.65. These initial transition layers both lie in (local) troughs of the energy density function, with the latter being a much higher local minimum (of the energy density). The system evolves by using the SED function with $\alpha = 0.37$, and equilibrium is reached at $t = 57.03$. The final equilibrium $\theta$ profile shows that grain 2 and grain 3 combine to form grain 4 by elimination of the $\partial_x \theta=0.65$ transition layer, and the transition $\partial_x \theta=1.2$ is retained between grain 1 and grain 4 even though both the transition layers belong to the SED function troughs. The reason for this behavior is that the activation energy required to eliminate the transition layer with $\partial_x \theta=0.65$ is much less than the activation energy required to eliminate the transition layer with $\partial_x \theta=1.2$. From the final $\theta$ profile in \fig{Fig8a}, it is also evident that grain growth occurs by grain rotation. 
The evolution of the initial condition described in \fig{Fig8a} with the CED function and $\alpha=0.37$ produces an equilibrium $\theta$ profile similar to the SED function. However, the equilibrium is achieved faster as compared to the SED function. \fig{EnergyVsTime_Cspline_Cusp_Combined_gt_1p2_0p65_a2_0p37_Multiplewells} shows the evolution of energy as a function of time corresponding to the SED and the CED functions. The equilibrium energy of the system is higher for the CED function as compared to the SED function and is achieved in less time. These results suggest that the dynamics of the system evolution and equilibrium energy of the system are influenced by the shape of the energy density functions used.
\subsection{Evolution from the initial condition with an arbitrary, nonequilibrium, random distribution of orientations}
\label{sec:random_initial_condition}
\noindent A recrystallized polycrystalline material consists of large numbers of grains with random orientations. Recrystallization is followed by grain growth. This case study focuses on the study of microstructure evolution that resembles such grain growth in an idealized sense.

\begin{figure}%%[h]
    \centering
    \begin{subfigure}{0.49\textwidth}
        \centering
        \includegraphics[width=1\linewidth]{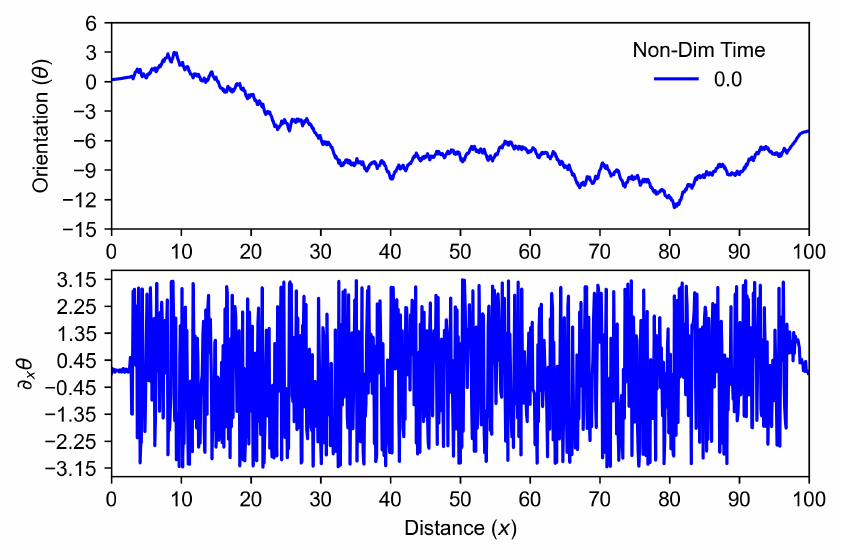}
        \caption{}
        \label{Fig9a}
    \end{subfigure}
    \hfill
    \begin{subfigure}{0.49\textwidth}
        \centering
        \includegraphics[width=1\linewidth]{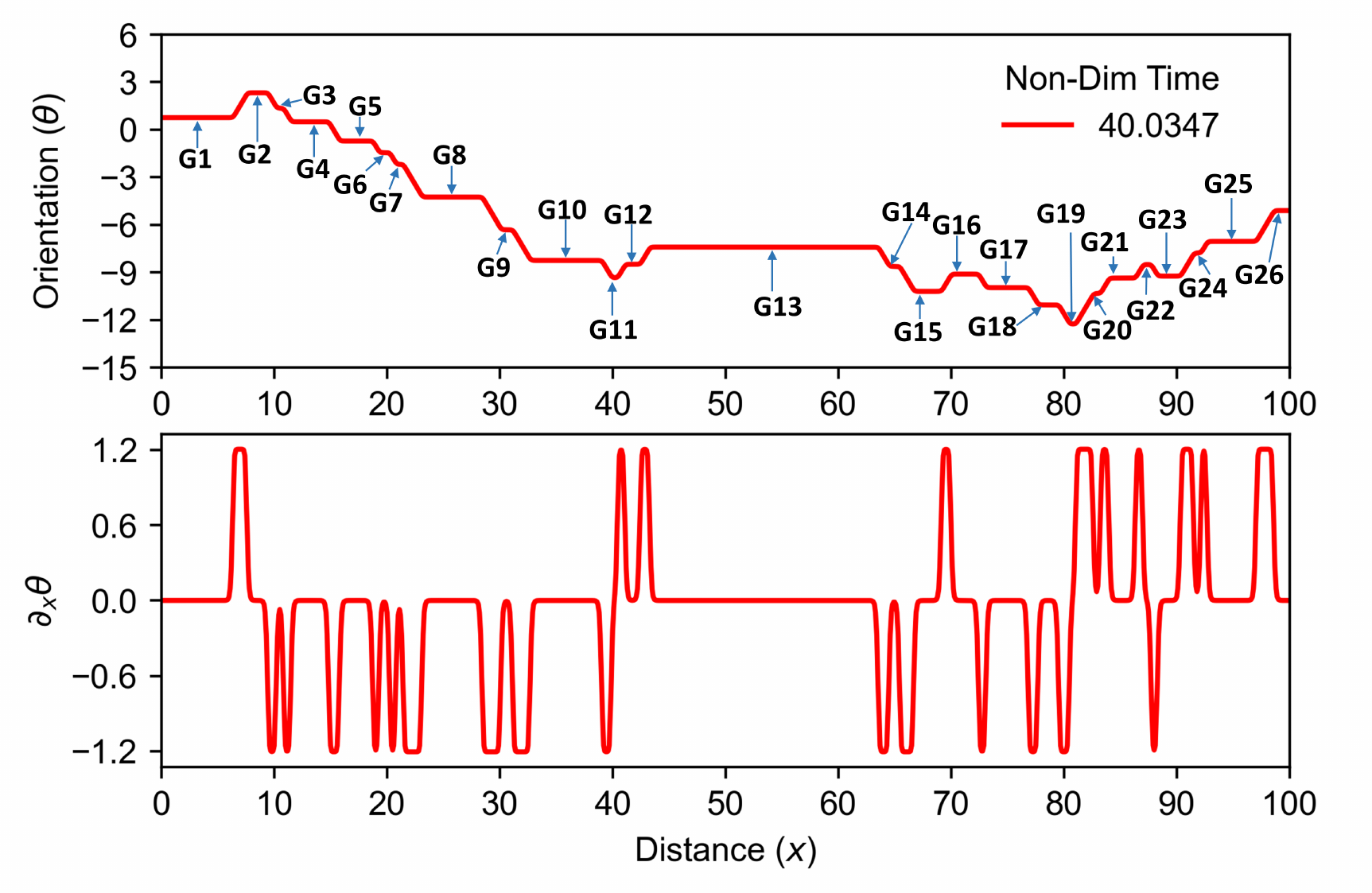}
        \caption{}
        \label{Fig9b}
    \end{subfigure}
    \hfill
    \begin{subfigure}{0.49\textwidth}
        \centering
        \includegraphics[width=1\linewidth]{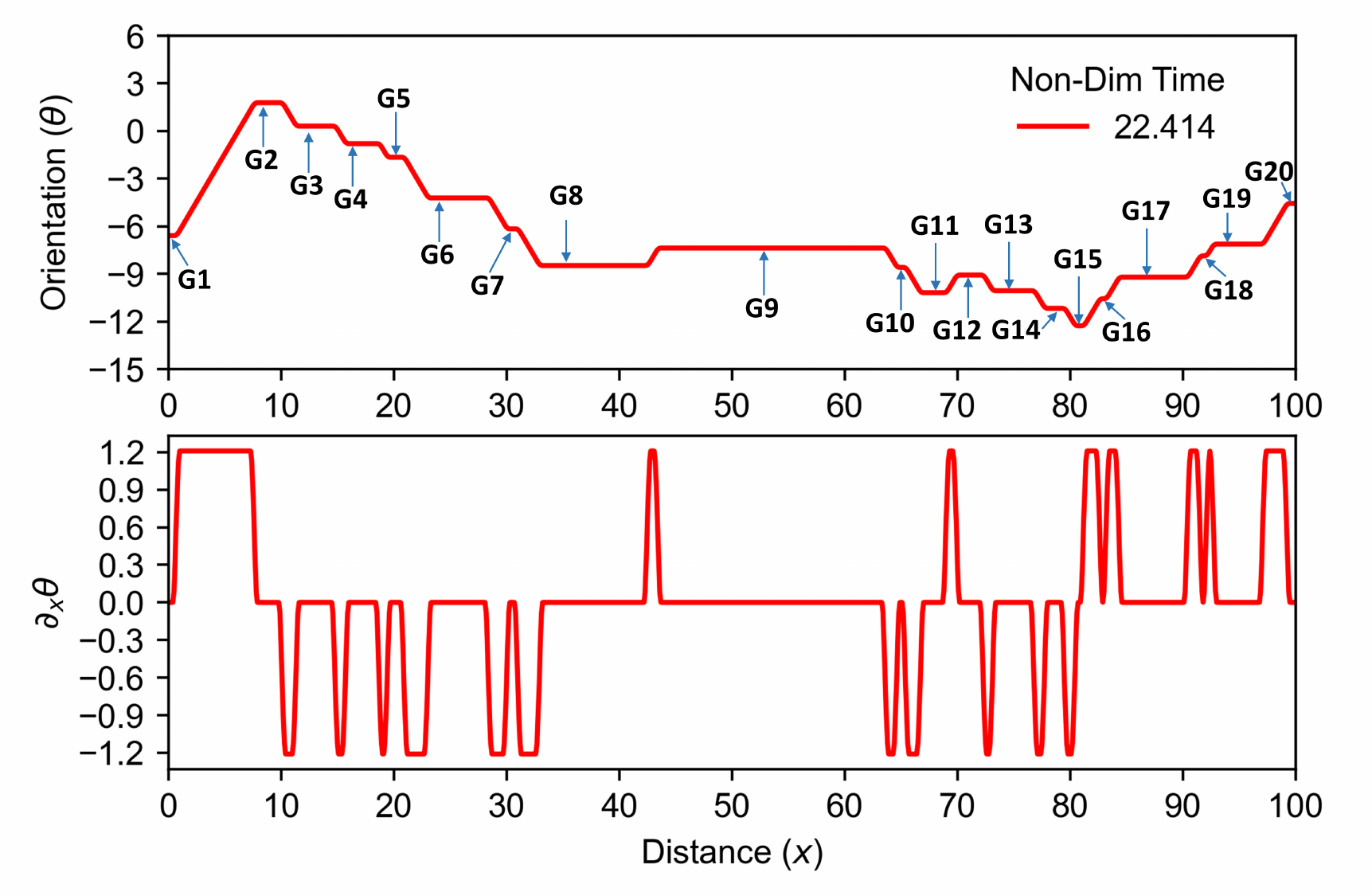}
        \caption{}
        \label{Fig9c}
    \end{subfigure}
    \hfill
    \begin{subfigure}{0.47\textwidth}
        \centering
        \includegraphics[width=1\linewidth]{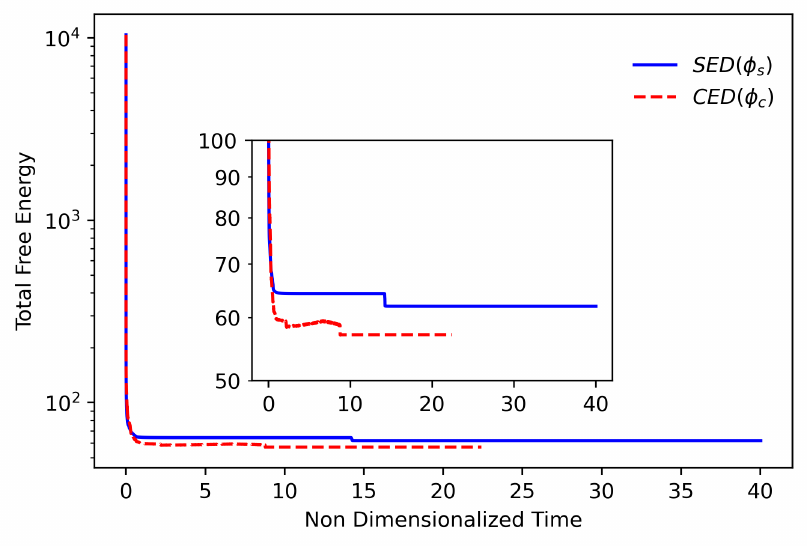}
        \caption{}
        \label{EnergyVsTime_Cspline_Cusp_Combined_RandomIC_E2_a2_0p37_Explicit}
    \end{subfigure}
    \caption{The temporal evolution of $\theta$ corresponding to initial random orientations as shown in (a) using the SED function (b) and using the CED function (c). The evolution of total free energy is shown in (d). The system evolves with $\alpha=0.37$ and produces the final equilibrium profile corresponding to $max|(\theta^{t+dt}-\theta^{t})/dt| < 1\times 10^{-3}$ and $|(F^{t+dt}-F^t)/dt| < 2 \times 10^{-4}$ where G1,G2,.,GN are the indices of the grains.}
\end{figure}

\fig{Fig9a} shows the initial condition with $\partial_x \theta$ varying randomly within the limit $-\pi$ to $\pi$. 
According to the definition of grain interior and grain boundary described in Section \ref{IC_single_TL}, there are 53 grains in \fig{Fig9a} and $\alpha$ is considered to be 0.37 for this simulation.

The initial condition shown in \fig{Fig9a} is set to evolve with both the SED and CED functions. The corresponding results are shown in \fig{Fig9b} and \fig{Fig9c} respectively. The final equilibrium $\theta$ profiles corresponding to the SED and CED functions have 26 and 20 grains, respectively.

The final equilibrium $\theta$ profiles in \fig{Fig9b} and \fig{Fig9c} also show that $\partial_x \theta=\pm1.2$, corresponding to the deepest local minima of the energy density, are the stable transition layers corresponding to $\alpha=0.37$ amongst all possible $\partial_x \theta$ available in the initial condition. The grain growth is higher for the CED function than for the SED function, even though the time taken by the CED function to achieve equilibrium is approximately half of the time taken by the SED function. The width of the grain boundary between grain 1 and grain 2, as shown in \mbox{\fig{Fig9c}}, after evolving with the CED function, is approximately equal to the size of the grains in the system. In general, physically, the width of the grain boundary is much smaller than the size of the grains. The wide transition layer may be considered as an idealized proxy of a grain in equilibrium containing a large number of subgrain boundaries in the form of dislocation walls, often observed after large deformations in real polycrystalline materials.

\fig{EnergyVsTime_Cspline_Cusp_Combined_RandomIC_E2_a2_0p37_Explicit} shows the evolution of energy with respect to the time corresponding to the SED and the CED functions. The energy of the final equilibrium $\theta$ profile is higher for the SED function than that for the CED function; the CED equilibrium result has fewer grain boundaries than the one that uses the SED function. 

\begin{figure}%%[h]
    \centering
    \begin{subfigure}{0.45\textwidth}
        \centering
        \includegraphics[width=1\linewidth]{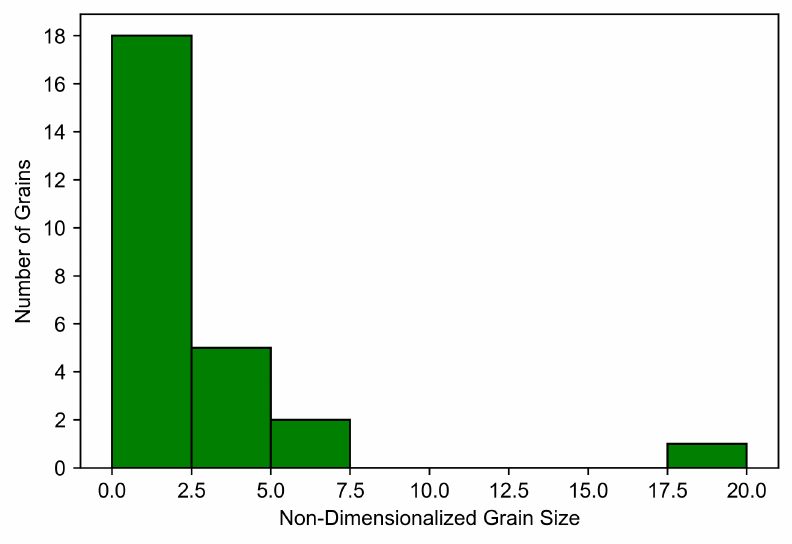}
        \caption{}
        \label{Fig10a}
    \end{subfigure}
    \hfill
    \begin{subfigure}{0.45\textwidth}
        \centering
        \includegraphics[width=1\linewidth]{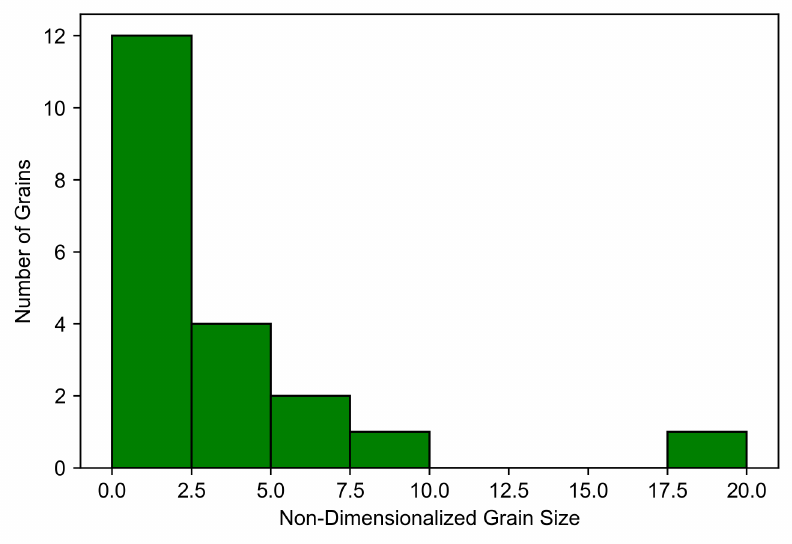}
        \caption{}
        \label{Fig10b}
    \end{subfigure}
    \caption{Number of grains in the final equilibrium $\theta$ profile using the SED function (a) as shown in \fig{Fig9b} and using the CED function (b) as shown in \fig{Fig9c}, binned based on grain size in each case.}
\end{figure}

The total number of grains decreases in both cases which signifies the occurrence of grain growth. The increased grain sizes are shown in \fig{Fig10a} and \fig{Fig10b}. The grains present in the equilibrium $\theta$ profile are binned according to their sizes in \fig{Fig10a} and \fig{Fig10b} corresponding to the SED and CED functions, respectively. A large fraction of grains has a non--dimensional size less than 10 in both the final equilibrium $\theta$ profile after evolution using the SED and CED functions. This signifies that both functions produce grains with almost similar sizes for the given initial condition. 

From these results, it is clear that the model is able to simulate idealized grain growth reminiscent of that in polycrystalline materials with a large number of randomly oriented `non-equilibrium' grains. The simulation also produces equilibrium grain boundaries in a 1--D grain array, including HAGBs, as is often observed after grain growth. As in earlier cases, the shape of the energy density function influences the dynamics of the evolution and the final equilibrium $\theta$ profile.
\subsection{Initial condition with a single, wide transition layer in a deep well}
\noindent This case study focuses on the stability of a wide transition layer that is an idealized representation of a grain with high dislocation content. Such grains are often observed after heavy, inhomogeneous deformation and have a continuously varying orientation \cite{HUMPHREYS200491,schwartz2009electron}.

The orientation profile in \fig{Fig9c} is used as an initial condition, and on evolution using the CED function shows that the wide $\partial_x \theta=1.2$ transition layer at the left of the domain seems to persist in time. The transition layer has a width close to the size of the grains present in the system. Therefore, this transition layer may be interpreted as a grain with a continuously varying orientation formed by more than one subgrain boundary coming close to each other as a result of the equilibration process.

\begin{figure}%%[H]
    \centering
    \begin{subfigure}{0.49\textwidth}
        \centering
        \includegraphics[width=1\linewidth]{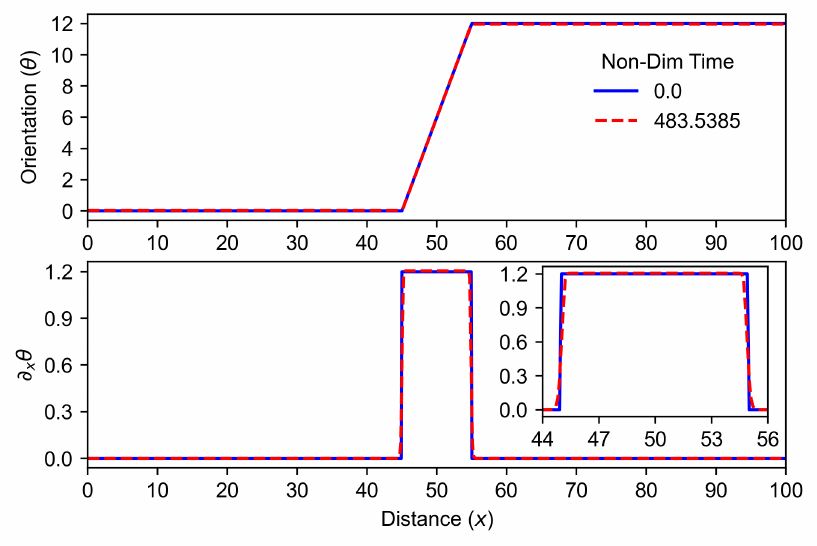}
        \caption{}
        \label{Fig11a}
    \end{subfigure}
    \hfill
    \begin{subfigure}{0.49\textwidth}
        \centering
        \includegraphics[width=1\linewidth]{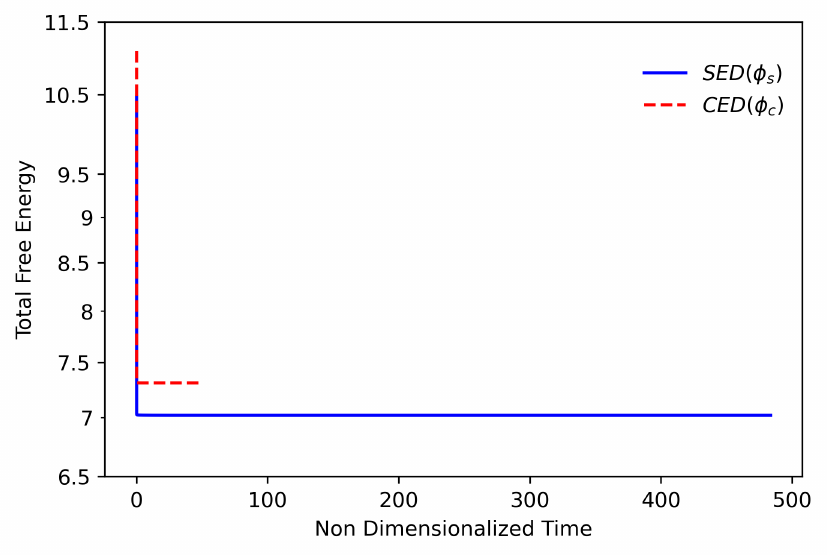}
        \caption{}
        \label{EnergyVsTime_Cspline_Cusp_Combined_WideGB10_gt_1p2_a2_0p37_Explicit}
    \end{subfigure}
    \caption{Temporal evolution of $\theta$ for initial conditions $\partial_x \theta$=1.2 wide transition layer ($l=10$) using the SED function, $\alpha=0.37$ is shown in (a) and the evolution of total free energy is shown in (b). The system produces the final equilibrium profile corresponding to $max|(\theta^{t+dt}-\theta^{t})/dt| < 1\times 10^{-14}$ and $|(F^{t+dt}-F^t)/dt| < 1 \times 10^{-14}$.}
\end{figure}

\fig{Fig11a} shows an initial condition with a wide transition layer placed between two grains and corresponding results, on evolution using the SED function. The system equilibrates satisfying $max|(\theta^{t+dt}-\theta^{t})/dt|<1\times 10^{-14}$ at $t = 483$. Evolution using the CED function shows results similar to those for the SED function.
\fig{EnergyVsTime_Cspline_Cusp_Combined_WideGB10_gt_1p2_a2_0p37_Explicit} shows the evolution of total energy of the system for both the SED and CED functions,  and as in most cases, the equilibrium $\theta$ profile produced by the CED function has higher energy than the equilibrium $\theta$ profile produced by the SED function and this is achieved sooner.
\begin{figure}%%[H]
    \centering
    \begin{subfigure}{0.49\textwidth}
        \centering
        \includegraphics[width=1\linewidth]{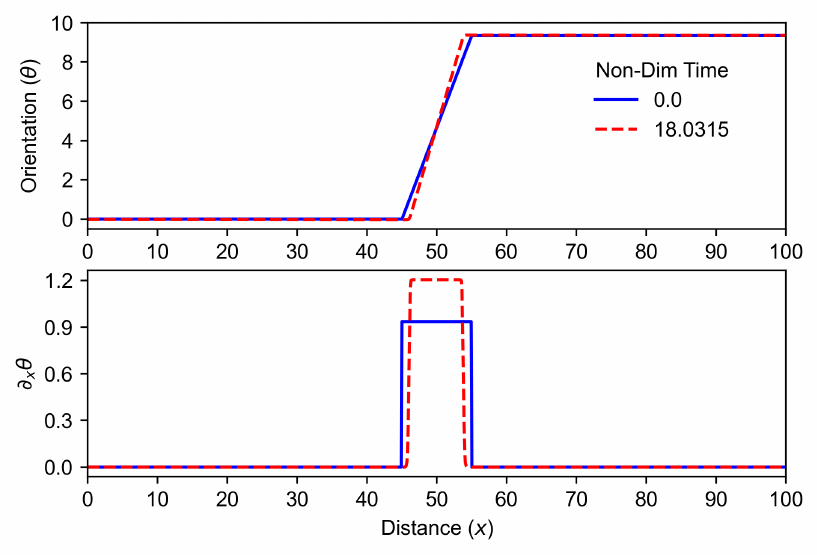}
        \caption{}
        \label{Fig12a}
    \end{subfigure}
    \hfill
    \begin{subfigure}{0.49\textwidth}
        \centering
        \includegraphics[width=1\linewidth]{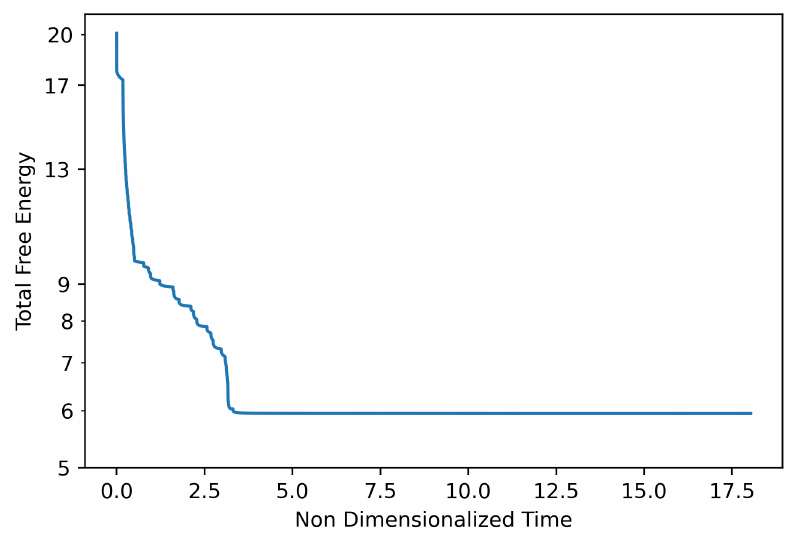}
        \caption{}
        \label{Fig12b}
    \end{subfigure}
    
\caption{Temporal evolution of $\theta$ in (a) $\partial_x \theta$=0.935 wide transition layer ($l=10$) initial conditions and corresponding total free energy is shown in (b). The energy density function is SED and $\alpha=0.37$. The system produces the final equilibrium profile corresponding to $max|(\theta^{t+dt}-\theta^{t})/dt| < 1\times 10^{-3}$ and $|(F^{t+dt}-F^t)/dt| < 5 \times 10^{-5}$.}
\end{figure}
Next, we explore the possibility of such a wide transition layer (equivalently, grains with high dislocation content) being produced from non--equilibrium initial condition (which can be interpreted as a partial signature of metastability of such configuration). A simulation is conducted with an initial wide transition layer situated away from the equilibrium trough at $\partial_x \theta = 1.2$. This is represented by initial condition $\partial_x \theta=0.935$ and is shown in \fig{Fig12a}, along with its evolution using the SED function. The width of the wide transition layer decreases to produce $\partial_x \theta=1.2$ transition layer. The total free energy decreases as shown in \fig{Fig12b}. 

\subsection{Handling kinks and non--convex energy density}\label{sec:kink_handling}
\noindent The initial condition in \fig{Fig7a} could evolve and produce a $\partial_x \theta=1.2$ equilibrium transition layer even with $\alpha=0$ as shown in \fig{Fig5a} since the orientation profiles at all times during evolution stay in the convex part of the energy density function, excluding the kinks which are handled by the jump condition. Thus, $\alpha>0$ is not necessary in this special case.

However, in general, situations arise naturally when non-convex regions of the energy density are visited by regions/subsets of the domain that are not single points.  In such situations, the evolution with $\alpha = 0$ is ill-posed, and the simulations produce non-physical oscillations in the  $\theta$ profile as shown in \ref{appendix:c}. This necessitates the use of $\alpha > 0$, which, incidentally, also arises naturally in the von-Neumann numerical stability analysis thresholds for our numerical scheme. We note that $\alpha > 0$ is not only eliminating the kinks but also plays a crucial role in making the $\theta$ profile stable to small amplitude, high wave-number perturbations as shown in \ref{appendix:a}.
\subsection{Grain Boundary Metastability}\label{gb_metastability}
\noindent

\begin{figure}%%[H]
    \centering
    \begin{subfigure}{0.45\textwidth}
        \centering
        \includegraphics[width=1\linewidth]{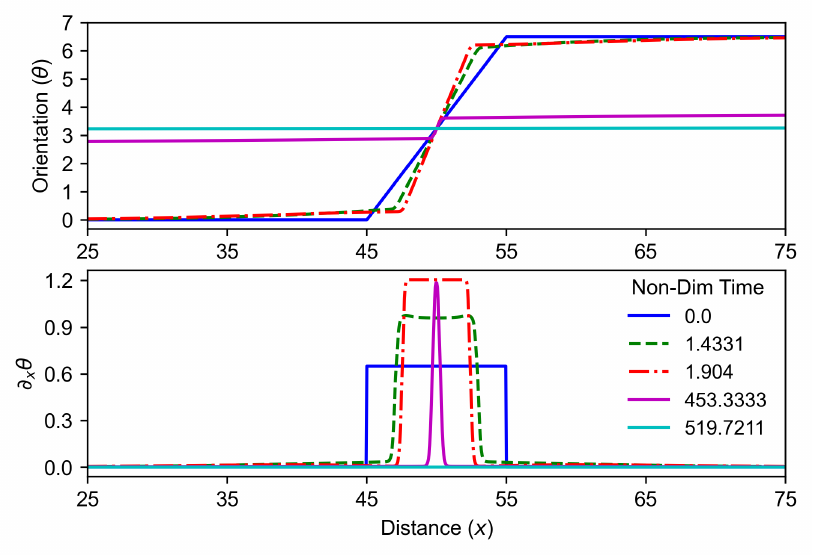}
        \caption{}
        \label{Fig13a}
    \end{subfigure}
    \hfill
    \begin{subfigure}{0.52\textwidth}
        \centering
        \includegraphics[width=1\linewidth]{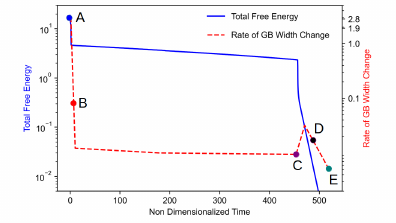}
        \caption{}
        \label{Energy_GBW_VsTime_widegb10_gt_0p65_a_0p37_dx_0p05_T_519p49}
    \end{subfigure}
    \caption{Temporal evolution of $\theta$ for initial condition $\partial_x \theta$=0.65 wide transition layer ($l=10$) using the SED function, $\alpha=0.37$, and mesh size ($\Delta x)$ = 0.05 is shown in (a). Figure (b) illustrates the evolution of total free energy (blue curve) and the rate of grain boundary width change (red curve). On the red curve, regions are labelled as follows. `A' to `B' corresponds to the transition from $\partial_x \theta = 0.65$ to $1.2$, `B' to `C' represents $\partial_x \theta = 1.2$, `C' to `D' denotes the transition from $\partial_x \theta = 1.2$ to $0$, and `D' to `E' represents the region where $\partial_x \theta \approx 0$. The system produces the final equilibrium profile corresponding to $max|(\theta^{t+dt}-\theta^{t})/dt| < 1\times 10^{-3}$ and $|(F^{t+dt}-F^t)/dt| < 5.1 \times 10^{-5}$.}
\end{figure}

\begin{figure}%%[H]
    \centering
    \begin{subfigure}{0.45\textwidth}
        \centering
        \includegraphics[width=1\linewidth]{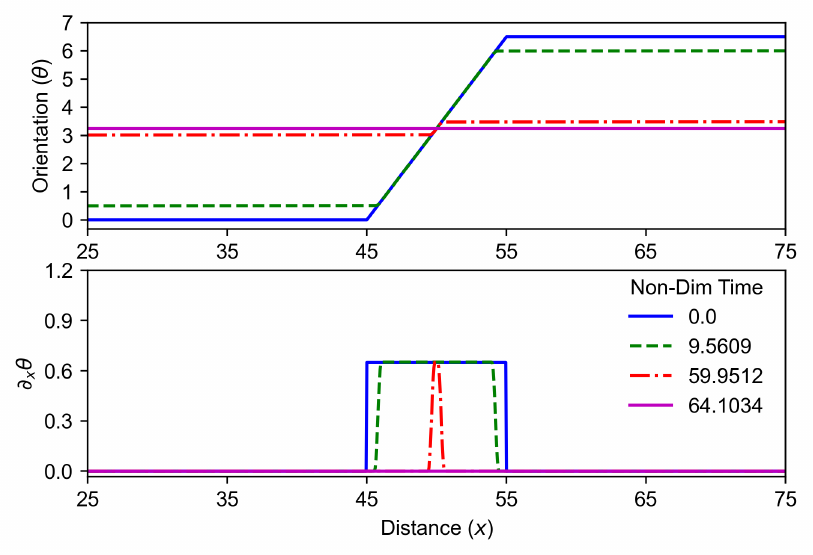}
        \caption{}
        \label{Fig14a}
    \end{subfigure}
    \hfill
    \begin{subfigure}{0.52\textwidth}
        \centering
        \includegraphics[width=1\linewidth]{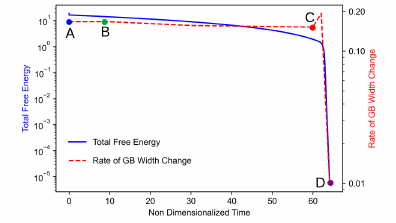}
        \caption{}
        \label{Energy_GBWVsX_CED_widegb10_gt_0p65_a_0p37_dx_0p05_T_64p1}
\end{subfigure}
\caption{Temporal evolution of $\theta$ for initial conditions $\partial_x \theta$=0.65 wide transition layer ($l=10$) using the CED function, $\alpha=0.37$, and mesh size ($\Delta x)$ = 0.05 is shown in (a). Figure (b) illustrates the evolution of total free energy (blue curve) and the rate of grain boundary width change (red curve). On the red curve, regions are labelled as follows. `A' to `B' represents $\partial_x \theta = 0.65$, `B' to `C' denotes the transition from $\partial_x \theta = 0.65$ to $0$, and `C' to `D' signifies the region where $\partial_x \theta \approx 0$. The system produces the final equilibrium profile corresponding to $max|(\theta^{t+dt}-\theta^{t})/dt| < 1\times 10^{-3}$ and $|(F^{t+dt}-F^t)/dt| < 5.1 \times 10^{-5}$.}
\end{figure}

\begin{figure}%%[H]
    \centering
    \begin{subfigure}{0.49\textwidth}
        \centering
        \includegraphics[width=1\linewidth]{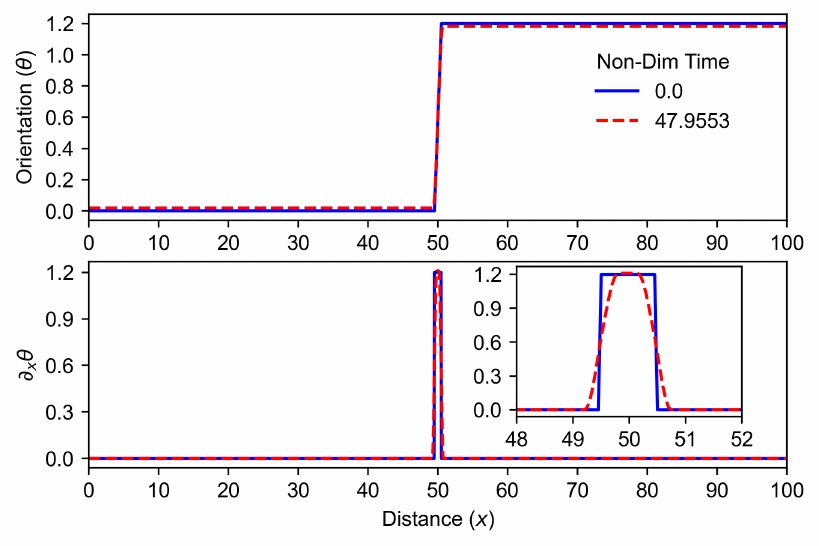}
        \caption{}
        \label{Fig15a}
    \end{subfigure}
    \hfill
    \begin{subfigure}{0.475\textwidth}
        \centering
        \includegraphics[width=1\linewidth]{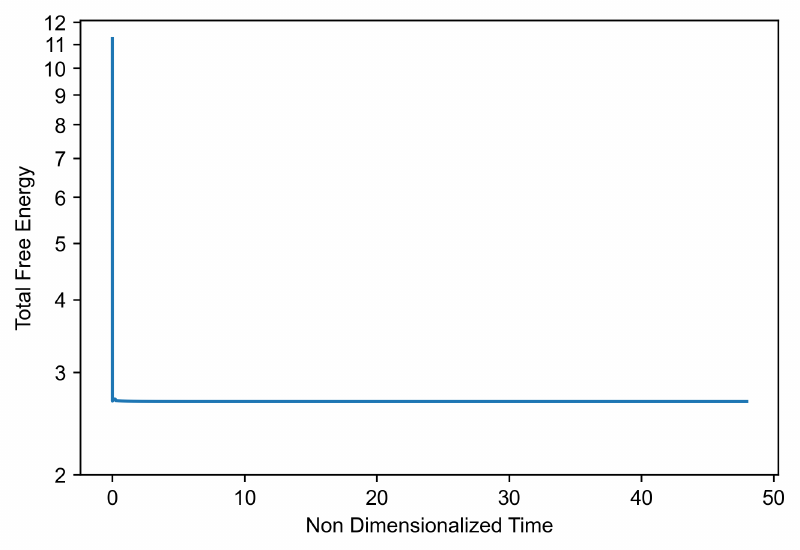}
        \caption{}
        \label{Fig15b}
    \end{subfigure}
    \caption{Temporal evolution of \mbox{$\theta$} is shown in (a) for \mbox{$\partial_x \theta=1.2$} narrow transition layer (\mbox{$l=1$})using the CED function for \mbox{$\alpha=0.37$} and a mesh size ($dx$) of 0.05. The total free energy evolution is shown in (b). The system produces the final equilibrium profile corresponding to \mbox{$max|(\theta^{t+dt}-\theta^{t})/dt|<1\times 10^{-8}$} and \mbox{$|(F^{t+dt}-F^t)/dt| < 1 \times 10^{-8}$}.}
\end{figure}

Experimental findings support the general thermodynamics belief that grain boundaries exist for a very long time in a metastable state and vanish in the limit of infinite time \cite{humphreys2004recrystallization}. This behavior is attributed to their preference for the lowest energy configuration due to the presence of fluctuations in energy. This case study aims to showcase the evolution of a few initial conditions with a temporally refined timestep and mesh size ($\Delta x =0.05$) to test the model's ability to capture this aspect over an extended period. \fig{Fig13a} shows an initial condition with a wide transition layer placed between two grains and corresponding results of evolution using the SED function. From \fig{Fig13a} and \fig{Energy_GBW_VsTime_widegb10_gt_0p65_a_0p37_dx_0p05_T_519p49}, the grain boundary with an initial $\partial_x \theta$ value of 0.65 quickly transitions to $\partial_x \theta = 1.2$ (within a non-dimensional time of $\sim 2$ marked with `A' to `B' in \fig{Energy_GBW_VsTime_widegb10_gt_0p65_a_0p37_dx_0p05_T_519p49}). It maintains this state for an extended period (approximately 453) while the grain boundary thickness gradually decreases at a low constant rate until it eventually vanishes ($\partial_x \theta \sim 0$). In \fig{Fig14a}, using the CED function with the same initial condition as \fig{Fig13a}, the grain boundary remains at $\partial_x \theta = 0.65$ for an extended period while the grain boundary thickness gradually decreases at a low constant rate before eventually vanishing ($\partial_x \theta \sim 0$). This prolonged retention at $\partial_x \theta = 0.65$ demonstrates the CED function's stronger metastability compared to the SED function. In \fig{Fig13a}, the grain boundary with $\partial_x \theta = 1.2$ using the SED function vanishes after a long time. However, in \fig{Fig15a}, the grain boundary with $\partial_x \theta = 1.2$ initial condition evolves using the CED function, and the system equilibrates at $t \sim 48$ with a very low residual ($max|(\theta^{t+dt}-\theta^{t})/dt|<1\times 10^{-8}$). Despite this, the $\partial_x \theta = 1.2$ grain boundary using the CED function survives, indicating stronger metastability than the SED function independent of the mobility. With a lower mobility term, grain boundaries evolving using the SED function can persist for extended periods, akin to those evolving with the CED function. The results demonstrate the model's capability to capture grain boundaries' metastable behavior, supporting the general thermodynamics belief that they exist for extended periods before eventually vanishing due to the lowest energy configuration favored by fluctuations over infinite time.
\section{Conclusions}\label{section:conclusion}
\noindent The following conclusions are derived from the present work:
\begin{itemize}
   
   \item A methodology is presented for incorporating experimentally measured grain boundary energy functions into a phase-field like formalism, and the work is placed in the context of existing literature, pointing to natural extensions for studying grain boundary network evolution in 3-D polycrystalline materials.
   \item The present model is demonstrated using two energy density functions, namely, a smooth energy density (SED) and a cusp energy density (CED) in 1--D. These are the commonly fitted functions to the experimentally measured grain boundary energy density data in the literature. The evolution dynamics of the CED function is faster than that of the SED function.
   \item A computational scheme is presented to handle the constraints arising from the presence of kinks in orientation profiles and the strong non-convexity of the experimentally measured energies.
   \item Various evolutions and their equilibria (in 1--D) recover idealized features of real physical systems such as equilibrium high--angle grain boundaries (HAGBs), grain rotation, grain growth, heavily deformed (static) microstructure often observed after the deformation process, and strong metastability. All the transition layers represent dislocation walls, and many of the equilibria resemble polygonized domains.
   \item The difference in the final equilibrium orientation profile, final energy of the equilibrium system, and evolution dynamics corresponding to different energy density functions for the same initial conditions and simulation parameters, suggest that the shapes of the energy density functions play a crucial role in the overall microstructure evolution.
\end{itemize}
While simplified, the 1–D situations our model addresses may be considered as a reasonable idealization of the state of moderately curved/straight grain boundaries away from junctions. Such ‘straight’ portions of grain boundaries of course move in reality, and it is a future challenge for our experimentally informed setup, possibly within a more encompassing description of coupled interfacial-bulk mechanics of defects as discussed in Section \ref{section:intro}, to demonstrate such behavior in higher dimensions as well as represent the mechanics of grain boundary junctions. This hope is supported by the observation of Glasner \cite{glasner2006grain} that, within the A--G/C--N models,  the ``dynamics of the grain boundaries arise from the curvature of the interfaces and variation in the “line energy” (associated with how sharply rolls meet at the grain boundaries) along each grain boundary'' (the latter interpreted as the energetics of disclination distributions within grain boundaries), as well as the wish expressed in \cite{glasner2006grain} expressed as ``more ambitious achievement would be to incorporate descriptions of disclinations and other one dimensional defects together with grain boundary dynamics.''
\section*{Acknowledgment}
The work of AA is supported by the NSF grant NSF OIA-DMR \#2021019. ALR acknowledges the financial support provided by IIT Bombay for Ph.D. student SA. ALR also acknowledges the partial financial support from Aeronautics Research Development Board (ARDB) vide Grant \#GTMAP1952 for part of a Computational Facility that was used for this work.
\section*{Conflict of Interest}
The authors have no relevant financial or non-financial interests to disclose.
\bibliographystyle{unsrtnat}
\bibliography{bibliography.bib}
%\printbibliography
%
%\clearpage
\appendix
\setcounter{figure}{0}
\section{Stability of the $\partial_x \theta = 1.2$ grain boundary/transition layer}
\label{appendix:a}
\begin{figure}[H]
    \centering
    \begin{subfigure}{0.49\textwidth}
        \centering
        \includegraphics[width=1\linewidth]{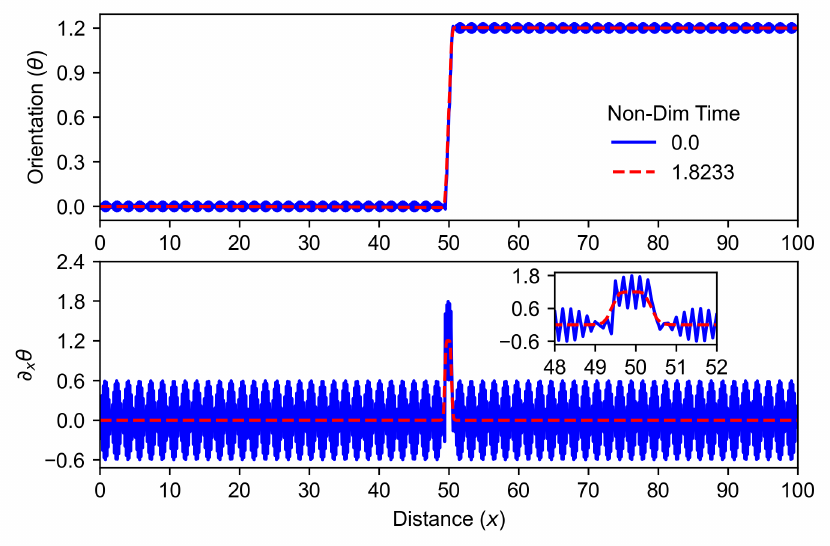}
        \caption{}
        \label{FigA1a}
    \end{subfigure}
%    \hfill
    \begin{subfigure}{0.475\textwidth}
        \centering
        \includegraphics[width=1\linewidth]{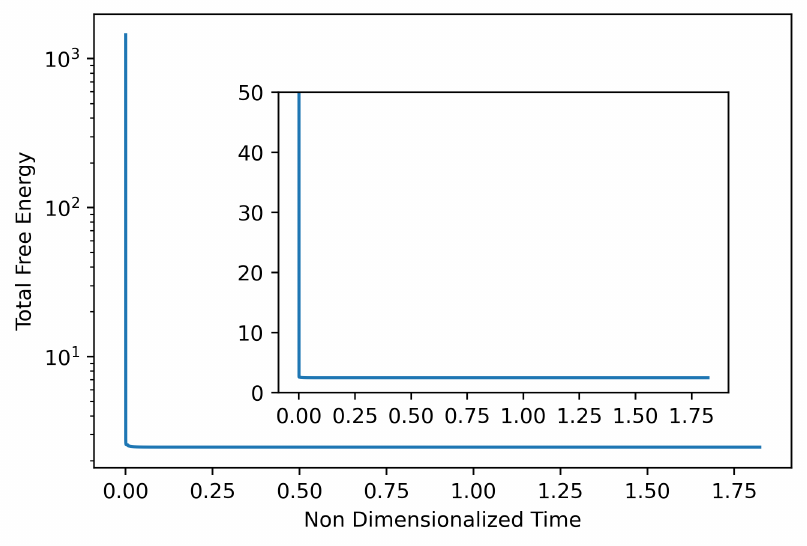}
        \caption{}
        \label{FigA1b}
    \end{subfigure}
    \caption{The temporal evolution of $\theta$ for the initial condition is shown in (a) $\partial_x \theta=1.2$ transition layer, with the addition of sine perturbations of the form $0.03\times \sin{\left(x/0.03\right)}$ and corresponding total free energy is shown in (b). The energy density function is SED and $\alpha=0.37$. The system produces the final equilibrium profile corresponding to $max|(\theta^{t+dt}-\theta^{t})/dt|<1\times 10^{-3}$ and $|(F^{t+dt}-F^t)/dt| < 2.9 \times 10^{-5}$.}
\end{figure}
\setcounter{figure}{0}
\section{Inadequacy of the $J_2$ `formulation'}
\label{appendix:b}
\begin{figure}[H]
    \centering
    \begin{subfigure}{0.49\textwidth}
        \centering
        \includegraphics[width=1\linewidth]{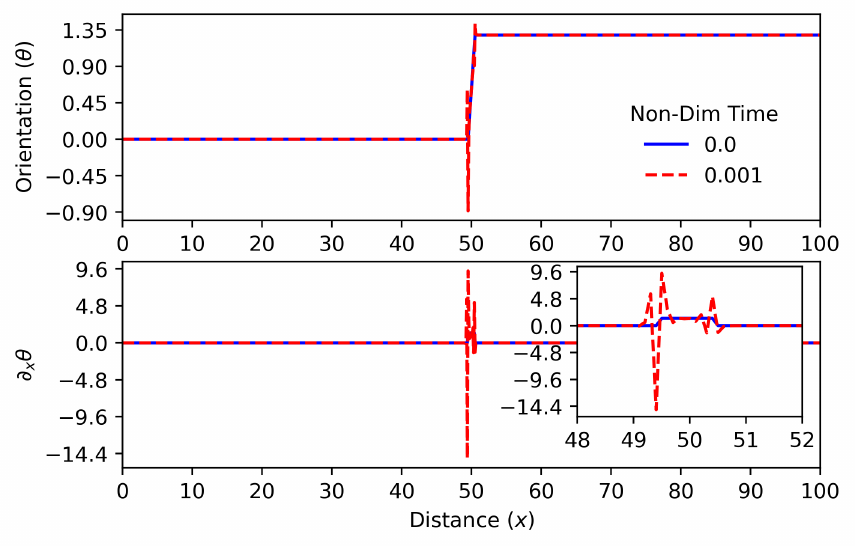}
        \caption{}
        \label{FigB1a}
    \end{subfigure}
%    \hfill
    \begin{subfigure}{0.475\textwidth}
        \centering
        \includegraphics[width=1\linewidth]{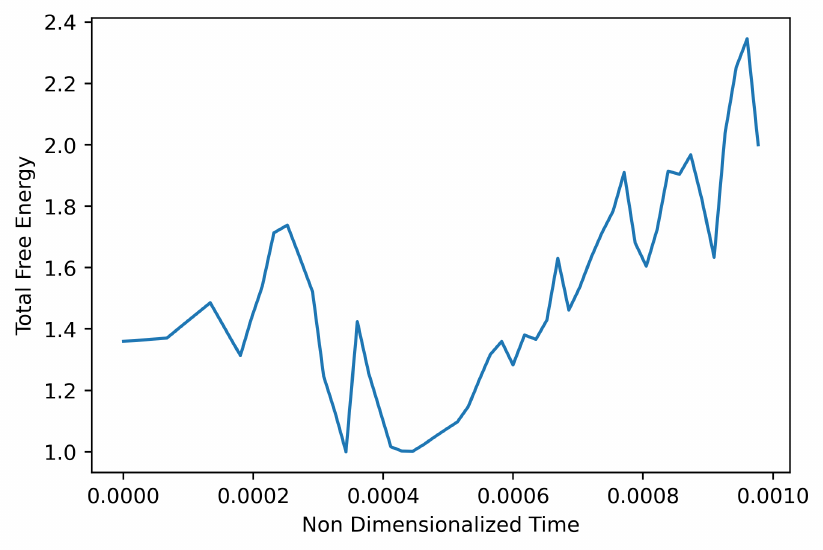}
        \caption{}
        \label{FigB1b}
    \end{subfigure}
    \caption{The temporal evolution of $\theta$ for the initial condition is shown in (a) $\partial_x \theta$=1.285 narrow transition layer ($l=1$) and the corresponding total free energy is shown in (b). The energy density function is SED and $\alpha=0$.}
    \label{J2_form_a2_0_gt_1p285}
\end{figure}
\begin{figure}%%[H]
    \centering
    \begin{subfigure}{0.49\textwidth}
        \centering
        \includegraphics[width=1\linewidth]{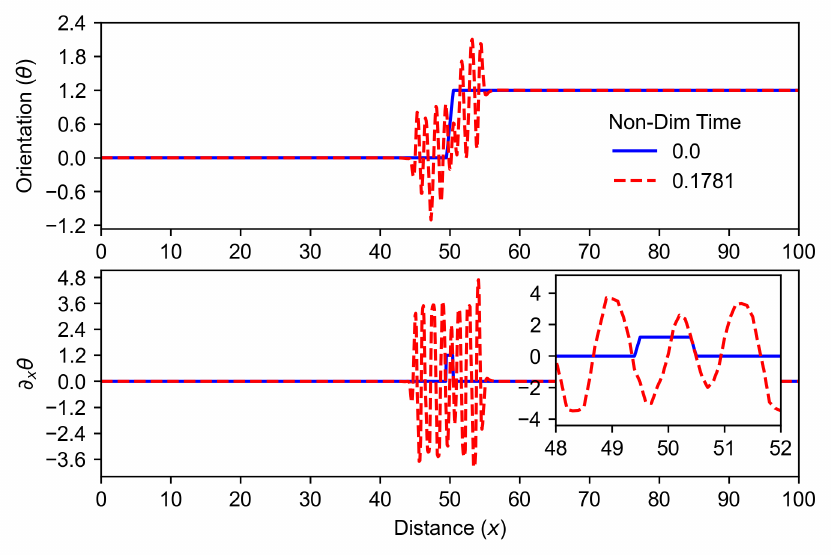}
        \caption{}
        \label{FigB2a}
    \end{subfigure}
    \hfill
    \begin{subfigure}{0.475\textwidth}
        \centering
        \includegraphics[width=1\linewidth]{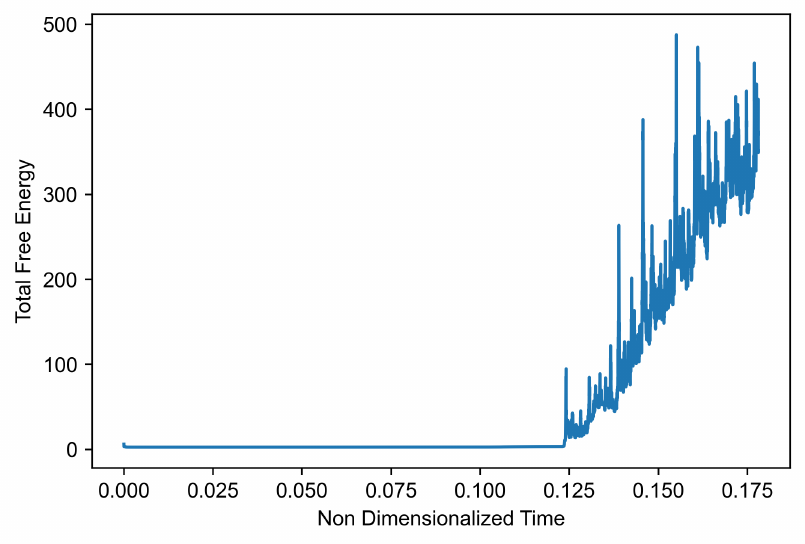}
        \caption{}
        \label{FigB2b}
    \end{subfigure}
    \caption{The temporal evolution of $\theta$ for the initial condition is shown in (a) $\partial_x \theta$=1.2 narrow transition layer ($l=1$) and the corresponding total free energy is shown in (b). The energy density function is SED and $\alpha=0.37$.}
\end{figure}
\begin{figure}[H]
    \centering
    \begin{subfigure}{0.49\textwidth}
        \centering
        \includegraphics[width=1\linewidth]{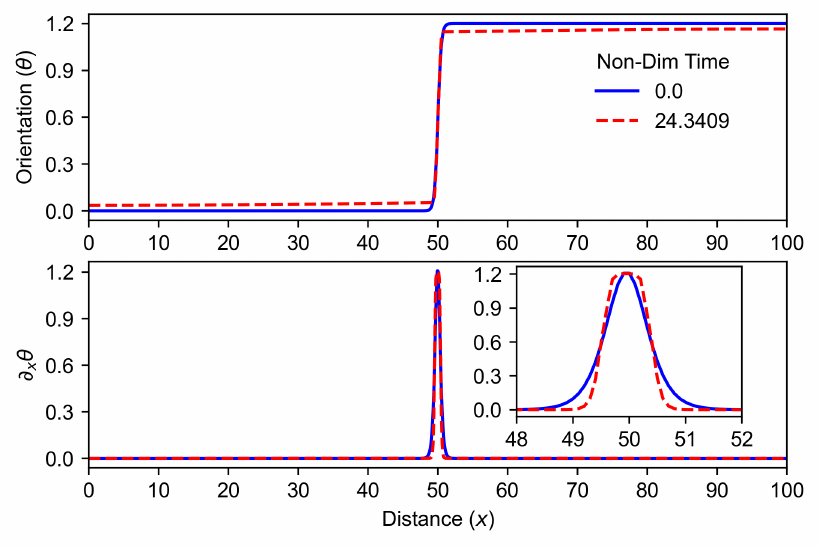}
        \caption{}
        \label{FigB3a}
    \end{subfigure}
    \hfill
    \begin{subfigure}{0.475\textwidth}
        \centering
        \includegraphics[width=1\linewidth]{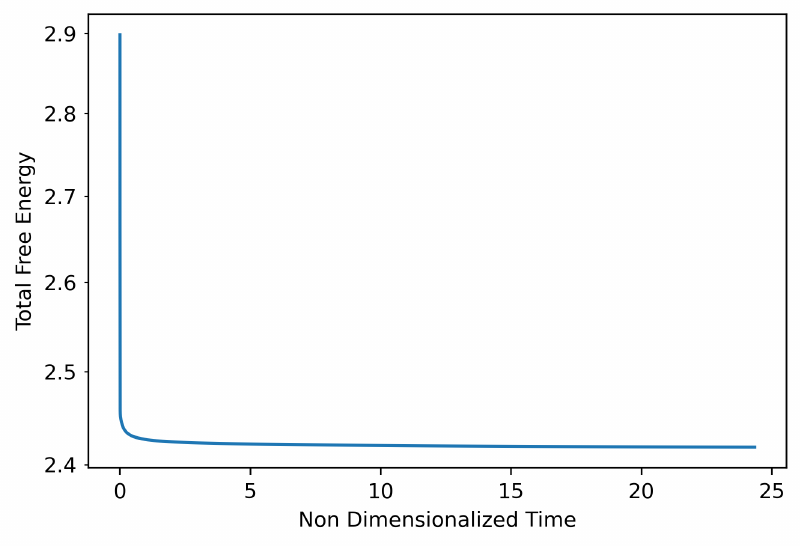}
        \caption{}
        \label{FigB3b}
    \end{subfigure}
    \hfill
    \begin{subfigure}{0.49\textwidth}
        \centering
        \includegraphics[width=1\linewidth]{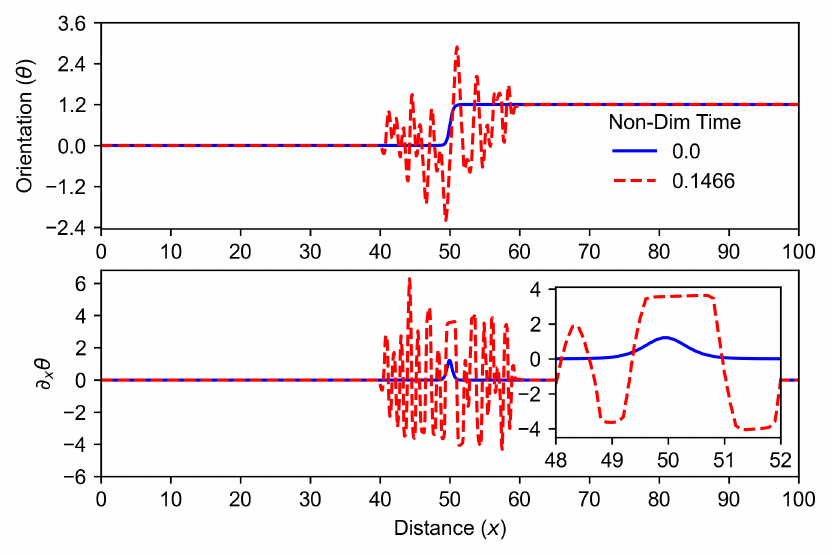}
        \caption{}
        \label{FigB3c}
    \end{subfigure}
    \hfill
    \begin{subfigure}{0.475\textwidth}
        \centering
        \includegraphics[width=1\linewidth]{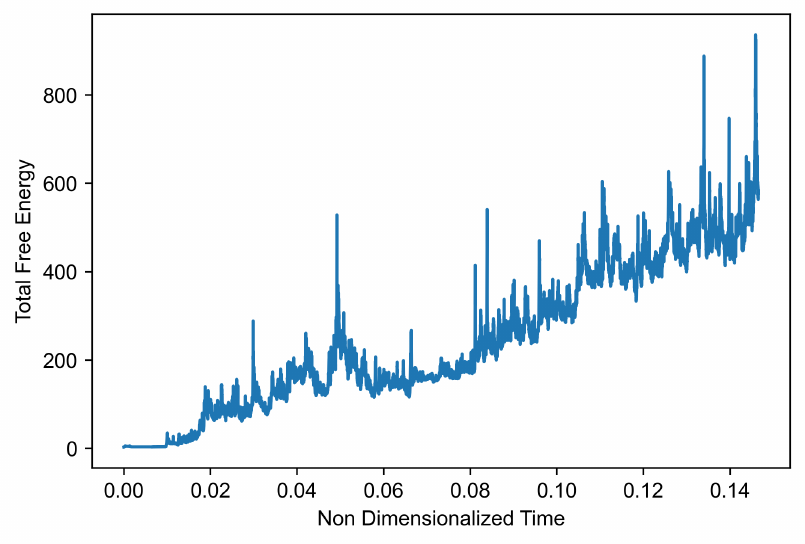}
        \caption{}
        \label{FigB3d}
    \end{subfigure}
    \caption{The figures in the left column show the temporal evolution of $\theta$ for smooth transition layer initial condition in (a) using $J$ formulation and (c) $J_2$ `formulation' corresponding to $\partial_x \theta$=1.2. The figures in the right column show the evolution of corresponding total free energies in (b) and (d), respectively. The energy density function is SED and $\alpha=0.37$.}
\end{figure}

\noindent The $J_2$ `formulation' in \eqn{EvolutionEqnJ2_2} is obtained by formally using the product rule on the $J$ term in \eqn{EvolutionEqn_Final_cspline}. The \mbox{\eqn{EvolutionEqnJ2_2}} is numerically discretized using the central difference formula in space and by the forward Euler method in time, and the corresponding fully explicit numerical discretization is given by
\begin{equation}\label{Numerical_Discretization J2form}
\frac{\theta_i ^{t+dt}-\theta_i ^{t}}{dt} = -\alpha \left[ {\frac{\theta_{i+2} ^t-4\theta_{i+1} ^t+6\theta_{i} ^t-4\theta_{i-1} ^t+\theta_{i-2} ^t}{(\Delta x)^4}}\right] + J_2\left({\partial_{x} \theta_{i}^t}\right)\left[{\frac{\theta_{i+1} ^t-2\theta_{i} ^t+\theta_{i-1} ^t}{(\Delta x)^2}}\right].
\end{equation}
\fig{FigB1a} shows the initial condition that is the same as the initial condition in \fig{Fig5e} and evolves using \eqn{EvolutionEqnJ2_2}. The initial condition $\partial_x \theta$=1.285 is slightly away from the equilibrium trough at $\partial_x \theta$=1.2 but it is still in the convex region of the SED function curve shown in \fig{GBEnergyVsGradTheta_Piecewise_Cubic_Splinefit}. The $\theta$ profile at $t = 0.001$ shows non-physical oscillations near the transition layer and the corresponding total free energy in \fig{FigB1b} also increases with oscillations. The system shows more non-physical behavior as time progresses. \fig{FigB2a} shows the initial condition that evolves using \eqn{EvolutionEqnJ2_2} with a small $\alpha$ ($\alpha$=0.37). The initial condition is the same as that of the initial condition in \fig{Fig7a}. Even with $\alpha=0.37$, the $\theta$ profile at $t = 0.1781$ shows oscillations near the transition layer, and the corresponding total free energy in \fig{FigB2b} also increases with oscillations similar to the initial condition in \fig{J2_form_a2_0_gt_1p285}. These results signify that \eqn{EvolutionEqnJ2_2} cannot handle the initial conditions with kinks whereas the same initial conditions in \fig{Fig5e} and \fig{Fig7a} evolve using \eqn{EvolutionEqn_Final_cspline} produce physically natural evolutions and the final equilibrium $\theta$ profiles. 
Even with a smooth initial transition layer, the $J_2$ `formulation'  produces non-physical $\theta$ profiles with progress in time as shown in \fig{FigB3c}. For the same initial condition, the $J$ formulation produces a physically reasonable $\theta$ evolution, as shown in \fig{FigB3a}.
\setcounter{figure}{0}
\section{Non-physical behavior for $\alpha=0$}
\label{appendix:c}
\begin{figure}[H]
    \centering
    \begin{subfigure}{0.49\textwidth}
        \centering
        \includegraphics[width=1\linewidth]{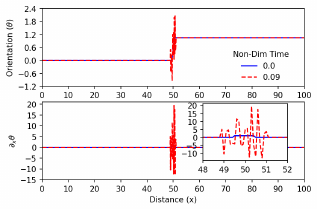}
        \caption{}
        \label{FigC1a}
    \end{subfigure}
    \hfill
    \begin{subfigure}{0.475\textwidth}
        \centering
        \includegraphics[width=1\linewidth]{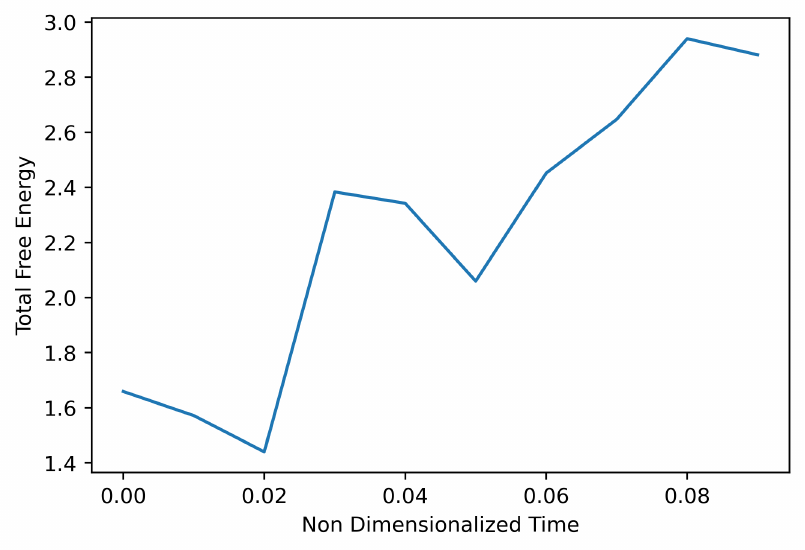}
        \caption{}
        \label{FigC1b}
    \end{subfigure}
    \caption{The temporal evolution of $\theta$ for the initial condition is shown in (a) $\partial_x \theta$=1.05 narrow transition layer ($l=1$) and the corresponding total free energy is shown in (b). The energy density function is SED and $\alpha=0$.}
\end{figure}

\noindent The initial condition $\partial_x \theta$=1.05 shown in \fig{FigC1a} is in the non--convex region of the SED function curve shown in \fig{GBEnergyVsGradTheta_Piecewise_Cubic_Splinefit} and $\alpha=0$. The $\theta$ profile at $t = 0.09$ shows non-physical oscillations near the transition layer and the corresponding total free energy in \fig{FigC1b} also increases with oscillations. The system shows more non-physical behavior as time progresses. Therefore, it is essential to introduce $\alpha>0$ to attain a physically reasonable $\theta$ evolution as demonstrated in Section \ref{sec:random_initial_condition}.
\end{document}